\documentclass[journal,twoside,web]{ieeecolor2}
\usepackage{generic}
\usepackage{cite}
\usepackage{amsmath,amssymb,amsfonts}
\usepackage{algorithmic}
\usepackage{graphicx}
\usepackage{textcomp}
\def\BibTeX{{\rm B\kern-.05em{\sc i\kern-.025em b}\kern-.08em
 T\kern-.1667em\lower.7ex\hbox{E}\kern-.125emX}}
\usepackage{mathtools}
\DeclarePairedDelimiter{\ceil}{\lceil}{\rceil}
\usepackage[utf8]{inputenc}
\usepackage[greek,english]{babel}
\usepackage{soul}
\usepackage{stfloats}
\usepackage{verbatim}
\begin{document}
\title{Control-Oriented Modeling and Layer-to-Layer Spatial Control of Powder Bed Fusion Processes}
\author{Xin Wang, Bumsoo Park, Robert G. Landers, Sandipan Mishra, Douglas A. Bristow
\thanks{X. Wang and D. Bristow are with Missouri University of Science and Technology, Rolla, MO 65409 (e-mails: {x.wang, dbristow}@mst.edu)}
\thanks{B. Park and S. Mishra are with Rensselaer Polytechnic Institute, 110 Eighth Street, Troy, NY USA 12180 (e-mails: {parkb5, mishrs2}@rpi.edu))}
\thanks{R. Landers is with University of Notre Dame, Notre Dame, IN 46556 (e-mail: rlanders@nd.edu)}
}

\maketitle

\begin{abstract}

Powder Bed Fusion (PBF) is an important Additive Manufacturing (AM) process that is seeing widespread utilization. However, due to inherent process variability, it is still very costly and time consuming to certify the process and the part. This has led researchers to conduct numerous studies in process modeling, in-situ monitoring and feedback control to better understand the PBF process and decrease variations, thereby making the process more repeatable. In this study, we develop a layer-to-layer, spatial, control-oriented thermal PBF model. This model enables a framework for capturing spatially-driven thermal effects and constructing layer-to-layer spatial controllers that do not suffer from inherent temporal delays. Further, this framework is amenable to voxel-level monitoring and characterization efforts. System output controllability is analyzed and output controllability conditions are determined. A spatial Iterative Learning Controller (ILC), constructed using the spatial modeling framework, is implemented in two experiments, one where the path and part geometry are layer-invariant and another where the path and part geometry change each layer. The results illustrate the ability of the controller to thermally regulate the entire part, even at corners that tend to overheat and even as the path and part geometry change each layer.
\end{abstract}


\section{Introduction}
\label{sec:introduction}

Powder Bed Fusion (PBF) is one of the most widely used Additive Manufacturing (AM) techniques for fabricating metal parts \cite{Singh}. This process uses a power source to melt and fuse powder into solid parts by scanning programmed regions of the powder bed layer by layer. While PBF can fabricate complex structures and small features, quality assurance is still a major challenge for commercial platforms \cite{McCann}. Therefore, significant research efforts have focused on monitoring and control of PBF processes. Various types of in-situ monitoring and ex-situ inspection sensors, such as acoustic \cite{Pandiyan}, optical \cite{Lott}, thermal \cite{Liu}, and computed tomography (CT) \cite{Lough2}, have been used to evaluate the quality of PBF parts.


For in-situ real-time sensing, optical measurements are frequently utilized for monitoring and control \cite{Renken2, Lott, Kanko, Cheng, Cheng2, Mazzoleni, Zhong}. For example, the Kruth group \cite{Clijsters} used an InfraRed (IR) camera and photodiodes to measure morphological features of melt pools (e.g., length, width, area) and radiation intensity for defect detection, and analyzed the process using time-domain data and spatial maps. Methods for closing the loop around any of the extracted features typically utilize real-time changes in the laser power as the process input. In doing so, these methods represent a standard temporal process control implementation. That is, while measurements may contain some spatial information, or be spatial variables themselves, they are acquired in real-time and control calculations are generated by temporal differential equations. Spatial effects of the underlying dynamics are implicitly treated as process disturbances.

\begin{figure*}[t]
\centerline{\includegraphics[width=7in]{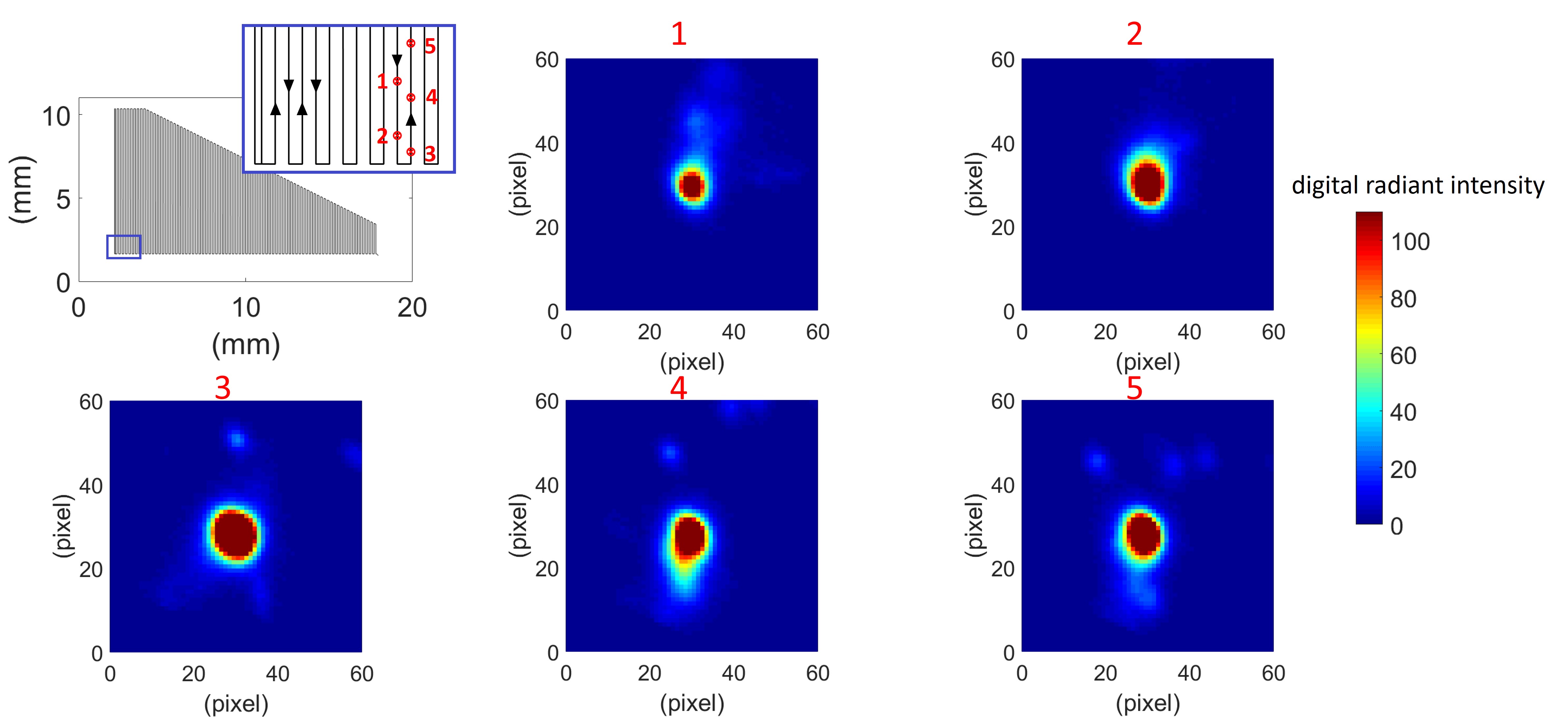}}
\caption{Example of Five Melt Pools Near a Turnaround Point}
\label{Fig_ThreeMelt1}
\end{figure*}

\begin{figure*}[t]
\centerline{\includegraphics[width=7in]{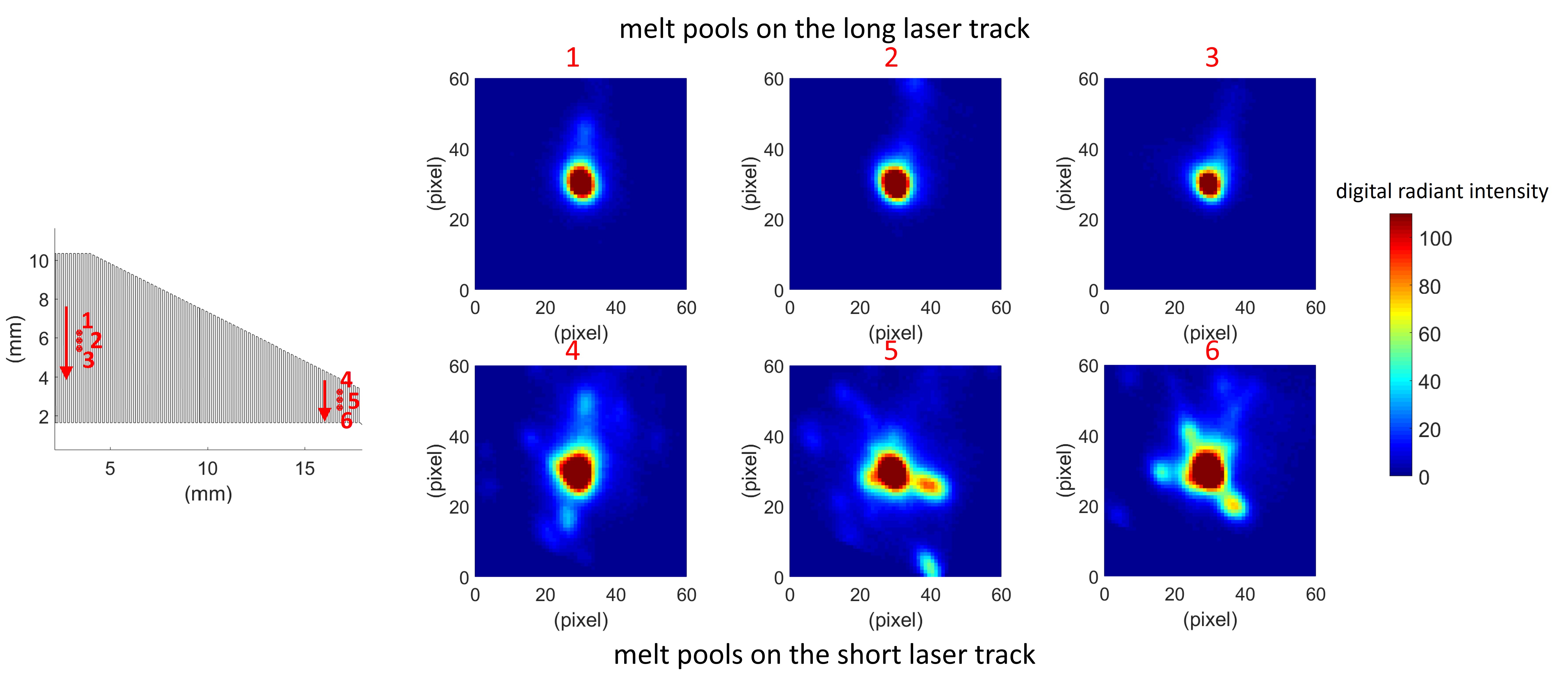}}
\caption{Example of Melt Pools on Long and Short Paths}
\label{Fig_ThreeMelt2}
\end{figure*}

Spatial dependence of melt pool dynamics can be seen in experiments. In open-loop experiments with constant laser power, two geometrical factors, the path and the boundary, influence melt pool size. Using the experimental set-up in Section V, some open-loop results are plotted in Fig.~\ref{Fig_ThreeMelt1} and Fig.~\ref{Fig_ThreeMelt2}. In Fig.~\ref{Fig_ThreeMelt1}, five successive melt pools near an edge are plotted. Because the thermal conductivity of powder is very low, the melt pool gets larger when it approaches the part border. The melt pools after turning around are larger than the melt pools before turning around because when the melt pool turns around the temperature of the previous laser track affects the new laser track. In Fig.~\ref{Fig_ThreeMelt2}, three melt pools on a long laser track on the left and a short laser track on the right are plotted. Because a shorter laser track allows shorter cooling time between laser tracks, the melt pool absorbs the energy of the previous track and gets hotter. These pictures illustrate two sources of non-uniformity of melt pool measurements in an open loop PBF process. 

Although implementation of temporal-domain control is straightforward, there are several critical disadvantages: 

\begin{itemize}
\item Many dynamic phenomenons in PBF, such as overheating around the corners of raster paths and variable cooling rates because of local part geometry differences, are spatially driven. These defects are difficult to incorporate in temporal analysis and control design.
\item Layer-to-layer dynamic analysis and control design are difficult to accomplish in temporal analysis because laser paths change with each layer. At the same time, useful information may be lost when previous layers having a similar geometry as the current layer are not incorporated into the control design. 
\item Thermal dynamics in PBF can be too fast for effective temporal feedback control because of the delays inherent in transporting and processing optical measurements from camera systems. 
\item Manufacturing researchers are more concerned about \textit{where} defects occur rather than \textit{when} they occur. 
\end{itemize}

\begin{figure*}[hb]
\centerline{\includegraphics[width=7in]{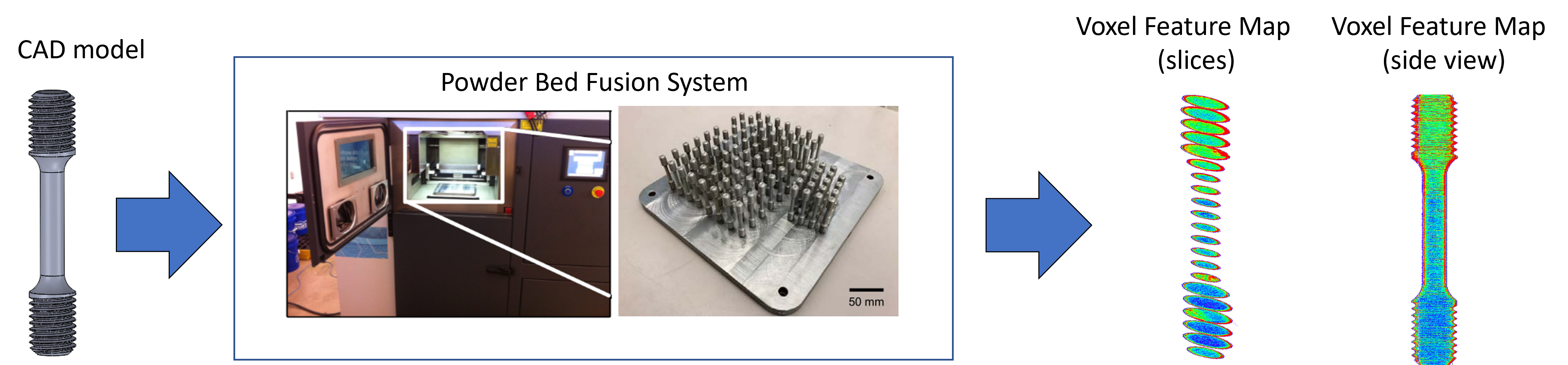}}
\caption{Powder Bed Fusion with Voxel Thermal Feature Map}
\label{Fig_CAMT}
\end{figure*}

As the last point above suggests, many manufacturing researchers have moved away from temporal monitoring. Instead, manufacturing researchers are increasingly using spatial feature maps, constructed from temporal data, in process monitoring. One such example uses the methods of the Krauss group \cite{Krauss, Krauss2} who defined a general form of location-based measurements (referred as ``key indicators"). Various real-time measurements can be fitted into the general form, such as thermal diffusivity, maximal temperature \cite{Krauss}, time above threshold and sputter activity count \cite{Krauss2}. These measurements are acquired in much the same way as the features used in temporal monitoring methods, but are then mapped to the spatial positions of the melt pools when they are acquired. The spatial position registration is performed layer by layer, making it possible to create 3D spatial feature maps. Figure ~\ref{Fig_CAMT} shows one such PBF manufacturing and monitoring system at the Missouri University of Science and Technology (MST), where an IR camera collects thermal data of a PBF process and then processes the data into a spatial feature map, or a so-called voxel feature map. The map features are then correlated with ex-situ spatial measurements from Computerized Tomography (CT) to aid in locating defects and optimizing processing parameters \cite{Lough2}. 

Layer-to-layer methods for spatially controlling other AM processes have been demonstrated. In the layer-to-layer methods, spatial measurements of a previous layer are acquired (potentially after the completion of that layer) and the control signal trajectory for the next layer is calculated offline between the layers. The Bristow and Landers group \cite{Sammons3, Gegel} modeled laser Direct Energy Deposition (DED) processes as a convolution operator in a spatial domain, and designed layer-to-layer controllers to stabilize deposition height in the repetitive process. The Barton and Hoelzle groups \cite{Hoelzle4, Hoelzle6, Hoelzle10} modeled an e-jet printing system by a 2D convolution formula where the height increment involves the convolution with spatial maps of droplet spreading behavior, and then they developed a spatial iterative learning controller. The Mishra group studied control of an ink-jet deposition process \cite{Guo, Inyang3} wherein the spatial convolution is learned by neural networks. However, unlike previous works on spatial control of deposition-type AM, the primary challenge in PBF is not the control of surface height, which is largely assured in PBF by the recoating process, but rather the size, temperature, and cooling rate of the melt pool. These dynamics are governed by the heat transfer process, which has a stronger temporal component than the pure deposition dynamics studied previously.

Some efforts to capture PBF dynamics in a spatial domain have been previously reported. The Rao group \cite{Yavari3} proposed a graph-theory based part-scale thermal model for Laser Powder Bed Fusion (LPBF). The term ``graph theory" refers to a strategy to choose random nodes in a part and represent the heat conduction path in any complex geometry. In experiments, a thermal camera and a photodiode were used to measure the temperature field. Although most of the effort concentrates on the temperature history instead of spatial control, the model provides a procedure to model the geometrical information in space topologically. The Mishra group previously proposed a control-oriented model for PBF with iterative learning control \cite{Spector}. They inspired and implemented the ideas of spatial control for PBF, but a universal spatial control theory for PBF still lies behind their work implicitly. A contribution of this paper is to formulate a voxel-based, formal mathematical spatial PBF model which lays a foundation for control analysis. 

This paper proposes a framework for spatial, layer-to-layer control of PBF. Such a framework provides several advantages over temporal methods:
\begin{itemize}
\item Spatially-driven dynamics, such as corner overheating and part geometry effects, can be explicitly incorporated.
\item The layer-to-layer control method reduces challenging delay and bandwidth requirements on sensors, as data only needs to arrive for control calculation before the start of the next layer.
\item Control is implemented on spatial data, in the same preferred manner as is currently being used for process monitoring by manufacturing researchers.
\end{itemize}

The remainder of the paper is organized as follows. Section II presents the spatial-temporal model of the PBF process. In Section III, the known laser trajectory is integrated into the spatial-temporal model to construct a purely spatial modeling framework that can be used in layer-to-layer control. The framework is used in Section IV to establish conditions on output controllability of the spatial voxel feature map using layer-to-layer control. A simple layer-to-layer control law is introduced in Section V and implemented experimentally. Concluding remarks are given in Section VI.

\section{Background: Time-Domain Dynamic Model}

Powder Bed Fusion is a complex physical process involving heat conduction, convection, radiation, phase transition and vaporization. Therefore, the melt pool microscale formation dynamics are complex. However, melt pool mesoscale geometry and thermal features can often be approximated from mesocale heat transfer calculations \cite{Zhang2} and used as a correlation proxy for important underlying mechanical properties \cite{Grasso}. Two simplifying assumptions are made here to the heat transfer model in PBF: 
\begin{itemize}
\item \textbf{Assumption 1}: The dominant heat transfer mode is heat conduction. 
\item \textbf{Assumption 2}: At the beginning of a new layer, the initial temperature is the ambient temperature. For convenience, the ambient temperature is set to be 0 by shifting the temperature field by an offset. 
\end{itemize}
The melt pool constitutes only a very small fraction of the part geometry and, therefore, may be considered as a point source in the heat transfer model. The convective heat transfer rate is about two orders of magnitude smaller than the conduction rate, and the radiative heat transfer in the melt pool is negligible because the size of the melt pool is negligibly small in comparison to the part \cite{H_Peng, R_Paul}, justifying Assumption 1. The second assumption is justified by noting that in most scenarios the cooling rate of metal parts in PBF is high enough for energy from the previous layer's processing to dissipate throughout the part and substrate during powder recoating between layers. Thus, any elevation in temperature state from ambient is well distributed and negligible. 

\begin{figure}[htbp]
\centering 
\includegraphics[width = 2.5in]{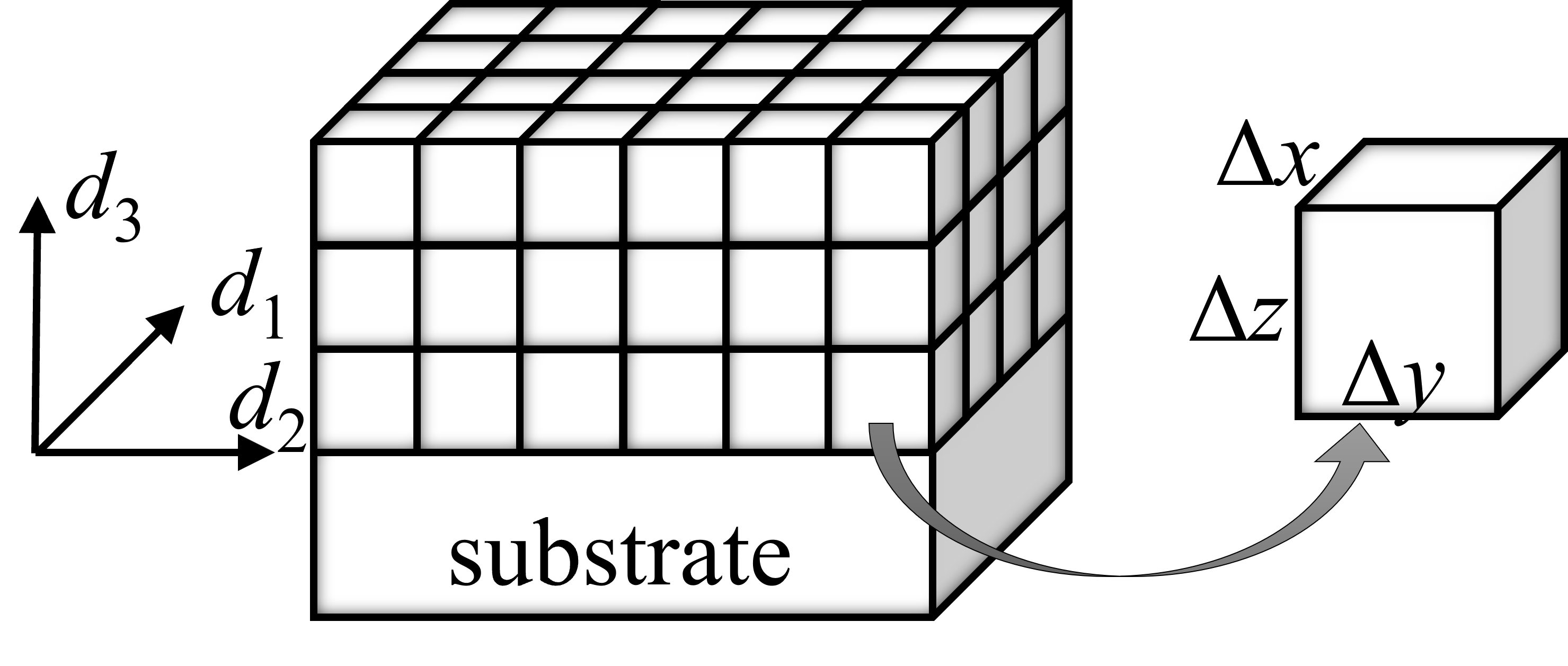}
\caption{Domain of Finite Difference Model}
\label{Fig_mesh}
\end{figure}

The governing equation of heat conduction is the Fourier-Biot equation 
\begin{equation} \label{Eq_Heat}
\Delta T+\frac{q}{k_\text{c}} = \frac{1}{\alpha}\frac{\partial T}{\partial t}, 
\end{equation}
\noindent where $\Delta$ is the Laplacian operator (m$^{-2}$), $T$ (K) is the temperature field, $q$ (W/m$^3$) is the heat power per volume, $k_\text{c}$ (W/(m·K)) is the conductivity, $t$ (s) is time and $\alpha$ (m$^2$/s) is the diffusivity. A finite difference form of the Fourier-Biot equation is 
\begin{equation} \label{Eq_Heat_FD}
\begin{split}
\frac{T_{d_1+1, d_2, d_3}(n)+T_{d_1-1, d_2, d_3}(n)-2T_{d_1, d_2, d_3}(n)}{\Delta x^2}
\\
+\frac{T_{d_1, d_2+1, d_3}(n)+T_{d_1, d_2-1, d_3}(n)-2T_{d_1, d_2, d_3}(n)}{\Delta y^2}
\\
+\frac{T_{d_1, d_2, d_3+1}(n)+T_{d_1, d_2, d_3-1}(n)-2T_{d_1, d_2, d_3}(n)}{\Delta z^2}
\\
+\frac{q_{d_1, d_2, d_3}(n)}{k_c}=\frac{1}{\alpha}\frac{T_{d_1, d_2, d3}(n+1)-T_{d_1, d_2, d3}(n)}{\Delta t}, 
\end{split}
\end{equation}
\noindent where $(d_1, d_2, d_3)$ is the 3D spatial index of an element, $n$ is the time index and $\Delta t$ (s) is the time step length. A rectangular domain is illustrated in Fig.~\ref{Fig_mesh}, where the size of an element is $\Delta x\times \Delta y\times \Delta z$, and a uniform temperature substrate boundary condition is illustrated. Adiabatic boundary conditions are used for the other surfaces \cite{H_Peng}. For convenience, $\Delta z$ is chosen as the layer thickness. 
To create a state-space model, the 3D temperature field needs to be vectorized into 1D. The vectorization operator is defined as follows: 

\vspace{6pt}
\noindent\textbf{Definition:} For any 3D array $\{a_{d_1, d_2, d_3}\}\in\mathbb{R}^{N_1\times N_2\times N_3}$ with the index ranges $d_i=1, 2, \cdots, N_i$ and $i=1,2,3$, the vectorization by the operator $\mathcal{V}:\mathbb{R}^{N_1\times N_2\times N_3} \rightarrow \mathbb{R}^{N_1N_2N_3}$ is 
\begin{equation} \label{Eq_VectorizeOp}
\mathcal{V}(\{a_{d_1, d_2, d_3}\})\triangleq\mathbf{a}\triangleq [\begin{matrix}a_{1} & \cdots & a_{\hat{d}} & \cdots & a_{N_1N_2N_3} \end{matrix}]^\text{T}, 
\end{equation}
\noindent where the vector index of $a_{\hat{d}}$ is
\begin{equation}\label{Eq_Vectorize}
\hat{d}=\phi(d_1, d_2, d_3)\triangleq d_1+(d_2-1)N_1+(d_3-1)N_1N_2. 
\end{equation}
\noindent The inverse operation of $\mathcal{V}$ is 
\begin{equation} \label{Eq_InvVectorizeOp}
\{a_{d_1, d_2, d_3}\}=\mathcal{V}^{-1}(\mathbf{a}), 
\end{equation}
\noindent where 
\begin{equation}\label{Eq_InvVectorize}
\begin{dcases}
d_3 = \ceil*{\frac{\hat{d}}{N_1N_2}}
\\
d_2 = \ceil*{\frac{\hat{d}-(d_3-1)N_1N_2}{N_1}}
\\
d_1 = \hat{d}-(d_2-1)N_1-(d_3-1)N_1N_2
\end{dcases}. 
\end{equation}

\begin{figure*}[htb]
\centering 
\includegraphics[width = 3.5in]{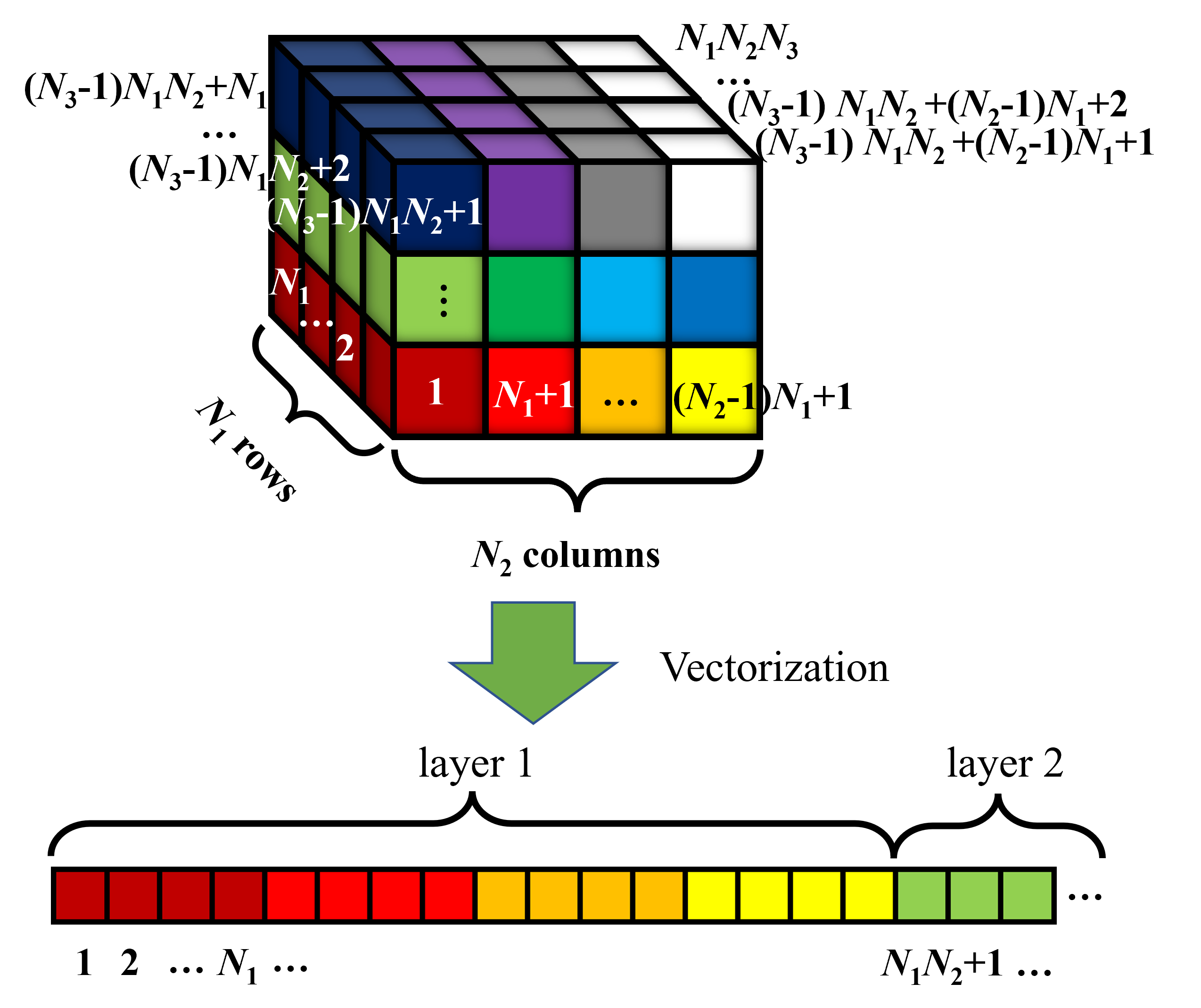}
\caption{Illustration of Vectorization Operator $\mathcal{V}$}
\label{Fig_vectorize}
\end{figure*}

\vspace{6pt}
\noindent The vectorization operator $\mathcal{V}$ is pictured in Fig.~\ref{Fig_vectorize}. 

\begin{figure*}[htb]
\centering 
\includegraphics[width = 4in]{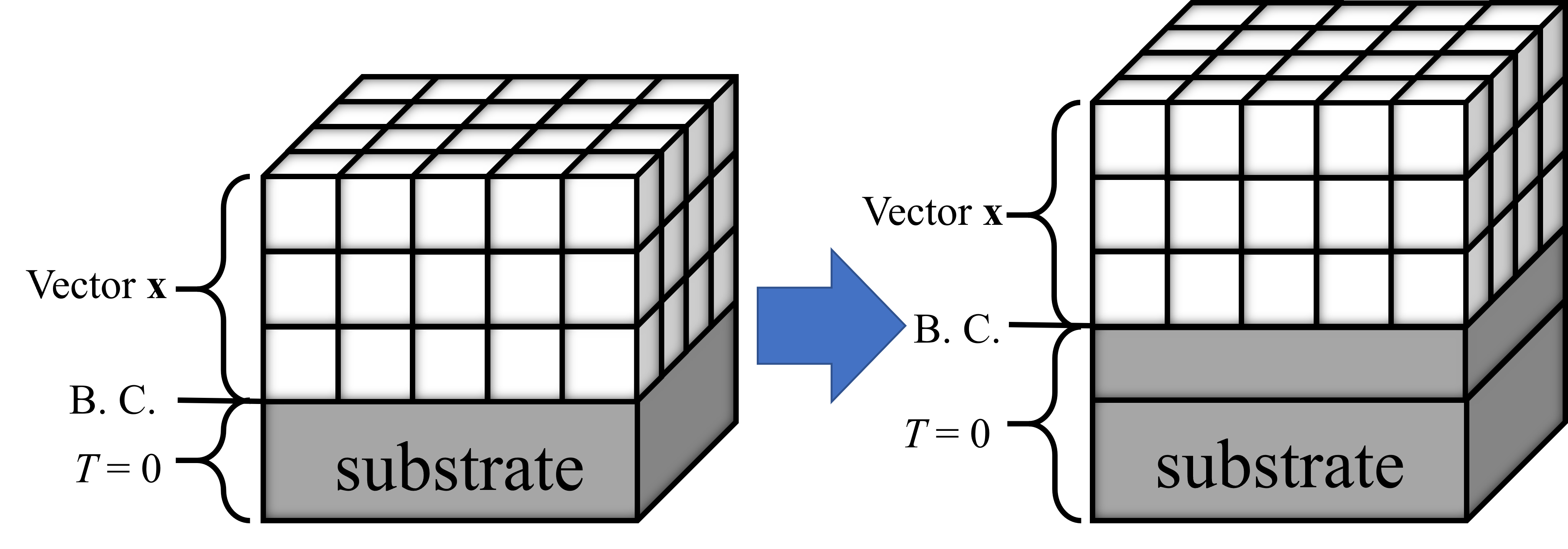}
\caption{Constant-Dimension State Vector $L=3$ and Boundary Condition (B. C.)}
\label{Fig_const}
\end{figure*}

To establish a control-oriented system equation, state-space variables are introduced. The temperature of an element is denoted by the state $x_{\phi(d_1, d_2, d_3)}$. For layer-to-layer control, a layer (i.e., iteration) number $l$ is added to the arguments so that $(n)$ is augmented to $(n, l)$. Then the vectorization from $T$ to $x$ is 
\begin{equation}\label{Eq_xDef}
x_{\phi(d_1, d_2, d_3)}(n, l)\triangleq T_{d_1, d_2, d_3}(n, l), 
\end{equation}
\noindent where $(n, l)$ is the $n$\textsuperscript{th} time step during the printing of the $l$\textsuperscript{th} layer. 

A practical problem is that the total number of elements in the state vector increases as new layers are added. Therefore, the system dimension increases with $l$. To maintain a constant system dimension, the bottom layer is considered to be completely solidified with the same temperature as the substrate (i.e., it becomes part of the bottom boundary) when a new layer is added to the top of the part. This concept is illustrated in Fig.~\ref{Fig_const} where, for example, the number of layers in the state vector is $L=3$. In this way, the dynamic system being controlled remains a constant size during the entire build. Then, the state vector at the $n^\text{th}$ time step during the $l^\text{th}$ layer constitutes only the top $L$ layers, namely, 
\begin{equation}\label{Eq_xStateDef}
\mathbf{x}(n, l)\triangleq \mathcal{V}(\{T_{d_1, d_2, d_3}(n, l)\}), 
\end{equation}
\noindent for $d_3 = l-(L-1), l-(L-2), ..., l$. Without specification, $d_1=1, 2, 3, ..., N_1$ and $d_2=1, 2, 3, ..., N_2$. 

The input power per volume dimensionalized into K is 
\begin{equation}\label{Eq_uDef}
u_{\phi(d_1, d_2, l)}(n ,l)\triangleq \frac{\alpha\Delta t}{k_\text{c}}q_{d_1, d_2, l}(n, l), 
\end{equation}
\noindent where the input power $q$ distributed over the surface space $(d_1, d_2, l)$ is transformed into a 1D vector space. The system input vector is
\begin{equation}\label{Eq_uVecDef}
\mathbf{u}(n, l)\triangleq \frac{\alpha\Delta t}{k_\text{c}}\mathcal{V}(\{q_{d_1, d_2, l}(n, l)\}). 
\end{equation}
\textbf{Remark}: the input vector $\mathbf{u}$ contains only the top elements (i.e., $N_1\times N_2$ elements in layer $l$) because the laser is only imposed on the surface. However, the temperature vector $\mathbf{x}$ contains the states of all $L$ layers (i.e., $N_1\times N_2\times L$ elements), because the states of the underlying layers, although they are not measured, are necessary to describe the heat transfer dynamics of the entire part. 

Based on the definitions of the input vector and the state vector, the heat conduction equation (\ref{Eq_Heat_FD}) applied to the domain under consideration yields 
\begin{equation}\label{Eq_system}
\mathbf{x}(n, l)=\mathbf{A}(l)\mathbf{x}(n-1, l)+\mathbf{B}(l)\mathbf{u}(n-1, l), 
\end{equation}
\noindent where $\mathbf{A}$ and $\mathbf{B}$ are derived from (\ref{Eq_Heat_FD}) and aligned by vectorization (\ref{Eq_VectorizeOp}). The matrix $\mathbf{A}$ describes the heat conduction physics and the geometrical connection of the elements in the governing partial differential equation, and $\mathbf{B}$ describes the instantaneous response of the temperature field resulting from $\mathbf{u}$. 

In (\ref{Eq_Heat_FD}),(\ref{Eq_system}), $q$ and $\mathbf{u}$ capture a general spatially distributed heat source. However, in PBF, the input vector takes a specific distribution and temporal path as defined by the slicer for that layer. Consider a laser power temporal signal $u_\text{t}(n, l)$ and a mask vector $\mathbf{h}(n, l)$, so that 
\begin{equation}\label{Eq_ut2u}
\mathbf{u}(n, l)=\mathbf{h}(n, l)u_\text{t}(n, l), 
\end{equation}
\noindent where $\mathbf{h}$ dictates the power distribution of $u_\text{t}$. 

\textbf{Remark}: the mask vector $\mathbf{h}$ varies with $(n, l)$, dictated by the laser path. For example, $\mathbf{h}(0,l)=[\begin{matrix}1&0&0&\cdots&0\end{matrix}]^\text{T}$ indicates the laser strikes the $1^\text{st}$ element on the top layer at $n=0$, and $\mathbf{h}(1,l)=[\begin{matrix}0&1&0&\cdots&0\end{matrix}]^\text{T}$ indicates the laser strikes the $2^\text{nd}$ element on the top layer at $n=1$. In addition, if the laser power is distributed, for example, $\mathbf{h}(0, l) = [\begin{matrix}1/8, 3/4, 1/8, 0, 0, \cdots, 0\end{matrix}]^\text{T}$ indicates the laser strikes elements 1, 2 and 3 at $n=0$ with 1/8, 3/4 and 1/8 of its magnitude, respectively. 

In a PBF system with optical sensing, the measurement is treated as a signal from the surface temperature field, namely,
\begin{equation}\label{Eq_Measure}
\begin{split}
y_\text{t}(n, l)=f_\text{M}(x_{N_1N_2(l-1)+1}(n, l), x_{N_1N_2(l-1)+2}(n, l),\\
\cdots, x_{N_1N_2l}(n, l)). 
\end{split}
\end{equation}
\noindent In general, $f_\text{M}$ can be any function which extracts some scalar features from an image of the top temperature field of the part. Define a linear sampling vector $\mathbf{f}$, which operates on the state vector $\mathbf{x}$, yielding 
\begin{equation}\label{Eq_MeasureLin}
y_\text{t}(n, l)=\mathbf{f}^\text{T}(n, l)\mathbf{x}(n, l). 
\end{equation}

\noindent\textbf{Example 1:} If the measurement is the summation of the surface temperatures (e.g., similar to a photodiode voltage \cite{Berumen}), then 
\begin{equation}\label{Eq_Ex1}
\mathbf{f}=\left[\begin{matrix}\mathbf{0}\\ \mathbf{1}\end{matrix}\right], 
\end{equation}
\noindent where $\mathbf{1}_{N_1N_2\times 1}$ is corresponding to the top layer and $\mathbf{0}_{N_1N_2(L-1)\times 1}$ is corresponding to the unmeasured underlying layers. 

\noindent\textbf{Example 2:} If the measurement is the maximal temperature \cite{Renken2,Krauss,Krauss2}, then 
\begin{equation}\label{Eq_MaxTemp}
y_\text{t}(n, l)=\max_{\hat{d}\in\Omega}\{x_{\hat{d}}(n, l)\}, 
\end{equation}
\noindent where $\Omega=\{N_1N_2(l-1)+1, N_1N_2(l-1)+2, \cdots, N_1N_2l\}$. It is reasonable to assume that the maximal temperature of an element occurs at the last moment it is struck by the laser, so the maximal temperature is approximately 
\begin{equation}\label{Eq_MaxTempAprx}
y_\text{t}(n, l)\approx x_{\hat{d}_\text{p}(n-1)}(n, l), 
\end{equation}
\noindent where $\hat{d}_\text{p}(n)$ is the laser position index at time index $n$. In this case 
\begin{equation}\label{Eq_Ex2}
\mathbf{f}(n, l)\approx[\begin{matrix}0&0&\cdots&0&1&0&\cdots&0\end{matrix}]^\text{T}.
\end{equation}
\noindent where $\mathbf{f}$ has a 1 at the position $\hat{d}_\text{p}(n-1)$ and 0’s elsewhere. 

\noindent\textbf{Example 3:} If the summation of melt pool temperatures is measured, then 
\begin{equation}\label{Eq_Ex3}
\mathbf{f}(n, l) = \left[\begin{matrix}f_1(n, l)\\ 
\vdots\\
f_{\hat{d}}(n, l)\\
\vdots\\
f_{N_1N_2L}(n, l)\end{matrix}\right], 
\end{equation}
\begin{equation}\label{Eq_Ex3bin}
\begin{cases}
f_{\hat{d}}(n, l)=1 & \begin{gathered}
\text{if } x_{\hat{d}+N_1N_2(l-L)}(n, l)\geq x_\text{th}\\
\text{and } \hat{d}>N_1N_2(L-1)
\end{gathered}
\\
f_{\hat{d}}(n, l)=0 & \text{otherwise}
\end{cases}, 
\end{equation}
\noindent where $x_\text{th}$ is the temperature threshold to determine if the pixel is part of the melt pool. 

The spatial-temporal PBF linear system model is written compactly as 
\begin{equation}\label{Eq_SyeyemComp}
\begin{cases}
\mathbf{x}(n, l)=\mathbf{A}(l)\mathbf{x}(n-1, l)+\mathbf{B}(l)\mathbf{p}(n-1, l)u_\text{t}(n-1, l)
\\
y_\text{t}(n, l)=\mathbf{f}^\text{T}(n, l)\mathbf{x}(n, l)
\end{cases}. 
\end{equation}

\section{Spatial Layer-to-Layer Model}
\begin{figure*}[h]
\centerline{\includegraphics[width=5in]{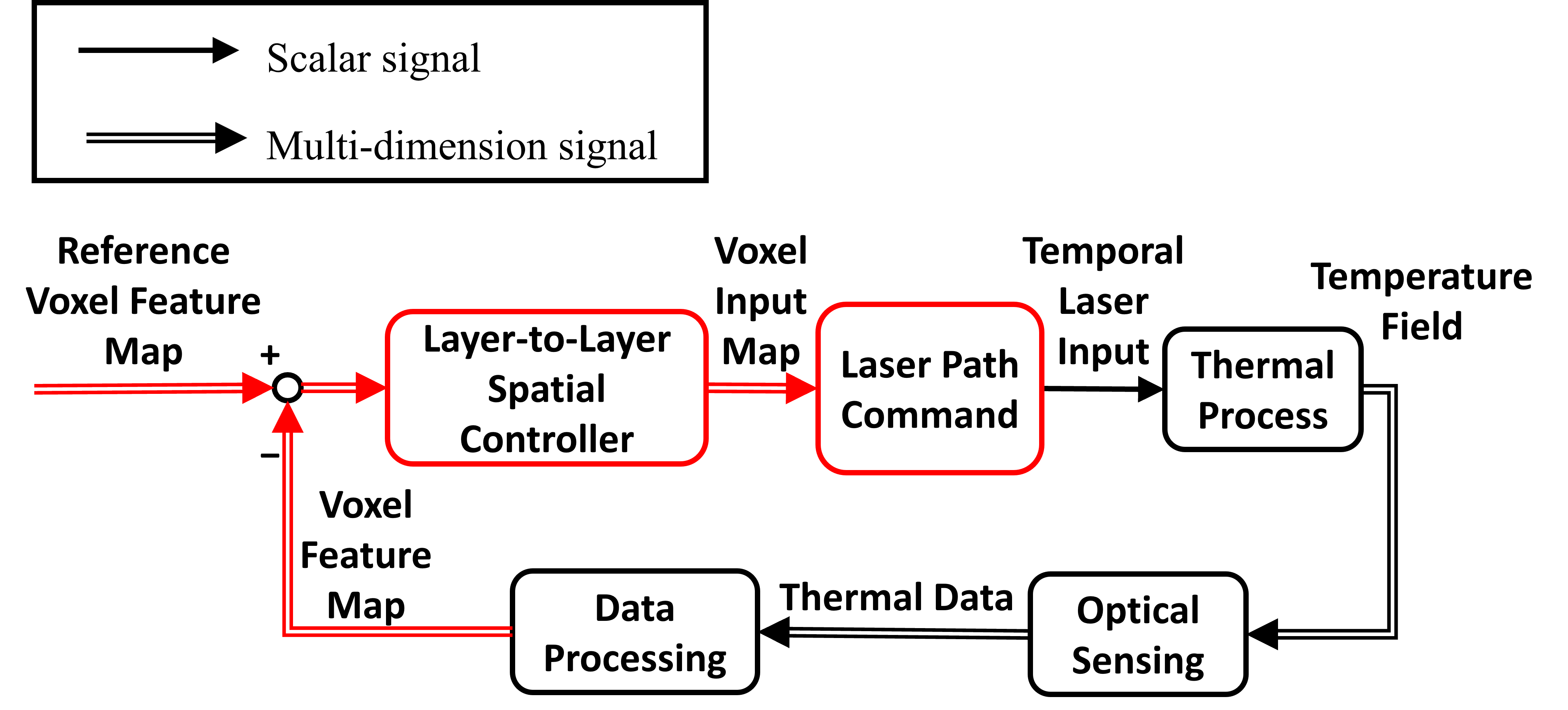}}
\caption{Layer-to-Layer Spatial Control Scheme for PBF}
\label{Fig_Flowchart}
\end{figure*}

As discussed in Section I, manufacturing researchers are more concerned about the
positions of defects rather than the time stamps of defects, so they have moved away from temporal monitoring to spatial feature monitoring. Melt pool features are acquired in much the same way as the features used in
temporal monitoring methods, but are mapped to the spatial positions of the melt pools. Spatial-domain layer-to-layer control is typically implemented as Fig.~\ref{Fig_Flowchart}, where the key idea of spatial control is marked by red. 

In the literature, it is sometimes unclear in which domain melt pool features are measured and, thus, the term ``melt pool feature" becomes technically ambiguous without a specific context. It is necessary to clarify the definitions of laser inputs and melt pool features both in the spatial domain and in the temporal domain, and to address the relationship between spatial and temporal variables. 

\begin{figure*}[htbp]
\centering 
\includegraphics[width = 5in]{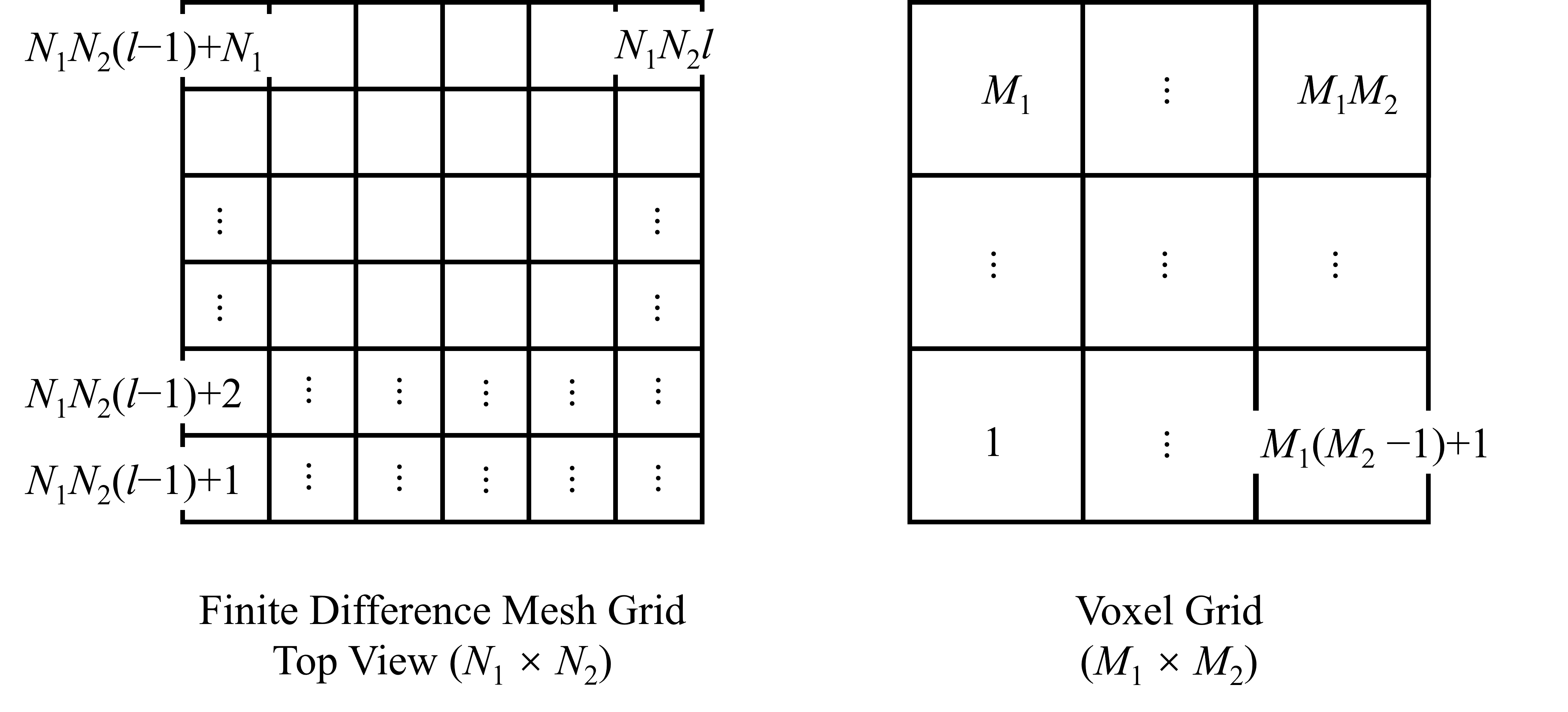}
\caption{Finite Difference Mesh Grid and Voxel Feature Grid}
\label{Fig_TwoGrids}
\end{figure*}

A voxel, a term from computer graphics, refers to an element on a regular grid in space, which is analogous with a pixel in a 2D image. Melt pool features mapped to a voxel grid constitute a ``voxel feature map". In a voxel feature map, the grid dimensions need not be the same as the meshing grid in the heat equation finite difference model. In a finite difference model, the meshing grid may be as fine as needed within the tolerance of computational cost. However, the voxel feature map grid size is restricted by the monitoring system. As is shown in Fig.~\ref{Fig_TwoGrids}, the two grids may have difference sizes, but their elements may be numbered in the same manner. 

Using the feature measurement function, (\ref{Eq_MeasureLin}), the temporal feature measurement is
\begin{equation}\label{Eq_ytVector}
\overline{\mathbf{y}}(l)\triangleq[\begin{matrix}y_\text{t}(1, l)&y_\text{t}(2, l)&\cdots&y_\text{t}(N_\text{t}, l)\end{matrix}]^\text{T}, 
\end{equation}
\noindent where $N_\text{t}$ is the last time instant of laser activity. Mapping the temporal measurement to the voxel feature grid, let $\mathcal{S}_i$ be the set of sample indices in voxel $i$ and $|\mathcal{S}_i|$ be the cardinality, which is the element number in a finite set, of $\mathcal{S}_i$. The voxel feature output vector
\begin{equation}\label{Eq_ysVector}
\mathbf{y}_\text{s}(l)\triangleq[\begin{matrix}y_{\text{s},1}(l)&\cdots&y_{\text{s},v}(l)&\cdots&y_{\text{s},N_\text{s}}(l)\end{matrix}]^\text{T}, 
\end{equation}
\noindent where $y_{\text{s}, v}$ is the voxel feature output at voxel $v$, is created from the temporal feature measurement vector by 
\begin{equation}\label{Eq_yt2ys}
\mathbf{y}_\text{s}(l)=\mathbf{Q}(l)\overline{\mathbf{y}}(l), 
\end{equation}
\noindent where the matrix $\mathbf{Q}$ is an operator that transforms the temporal sequence of measurements into a spatial vector. A natural example of $\mathbf{Q}$ is an average-within-the-voxel operator 
\begin{equation}\label{Eq_Q}
\mathbf{Q}=\{q_{ij}\}, 
\end{equation}
\begin{equation}\label{Eq_qij}
q_{ij}=\begin{cases}
1/|\mathcal{S}_i| & \text{if }j\in\mathcal{S}_i\\
0 & \text{otherwise}
\end{cases}. 
\end{equation}

\begin{figure*}[htb]
\centering 
\includegraphics[width = 6in]{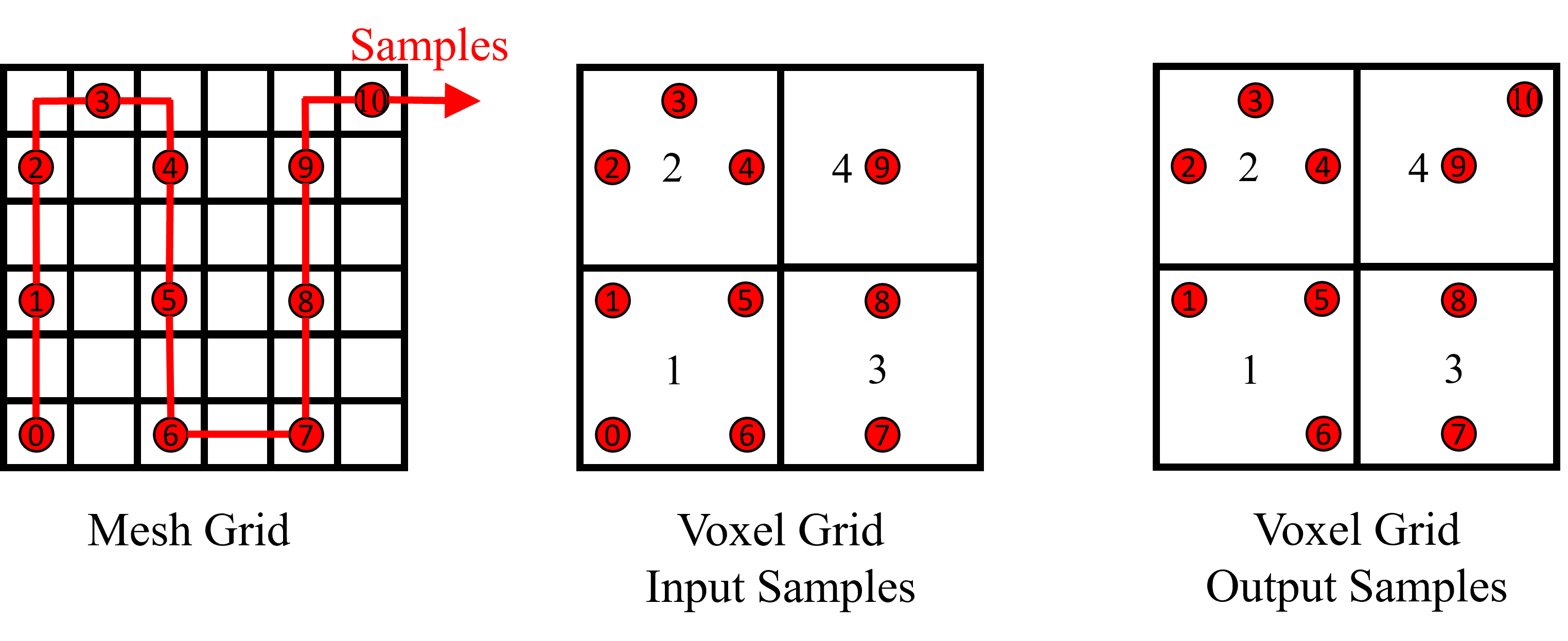}
\caption{Example of Meshing Grid of Top Layer and Spatial Control Grid}
\label{Fig_Ex4}
\end{figure*}

\noindent\textbf{Example 4:} Consider an example laser path as illustrated in Fig.~\ref{Fig_Ex4}, using any feature measurement function (\ref{Eq_MeasureLin}). As defined in (\ref{Eq_ytVector}), the temporal feature measurements span a temporal index range from 1 to 10, yielding temporal outputs, $y_\text{t}(1)$ to $y_\text{t}(10)$. The sampling rate over the finite difference mesh grid is 2 elements/sample with a hatch spacing of 2 elements. The voxel grid size is three times the meshing grid size. The sets of sample indices in the voxels are
\begin{equation}
\mathcal{S}_1=\{0, 1, 5, 6\}, 
\end{equation}
\begin{equation}
\mathcal{S}_2 = \{2, 3, 4\}, 
\end{equation}
\begin{equation}
\mathcal{S}_3 = \{7, 8\}, 
\end{equation}
\begin{equation}
\mathcal{S}_4 = \{9, 10\}, 
\end{equation}
In this example, the matrix $\mathbf{Q}$ for the average-within-the-voxel is 
\setcounter{MaxMatrixCols}{12}
\begin{equation}\label{Eq_ExQ}
\mathbf{Q}=
\begin{bmatrix}
\frac{1}{3} & 0 & 0 & 0 & \frac{1}{3} & \frac{1}{3} & 0 & 0 & 0 & 0 \\
0 & \frac{1}{3} & \frac{1}{3} & \frac{1}{3} & 0 & 0 & 0 & 0 & 0 & 0 \\
0 & 0 & 0 & 0 & 0 & 0 & \frac{1}{2} & \frac{1}{2} & 0 & 0 \\
0 & 0 & 0 & 0 & 0 & 0 & 0 & 0 & \frac{1}{2} & \frac{1}{2}
\end{bmatrix}. 
\end{equation}

Based on the same voxel grid with its numbering, the voxel input vector is defined in a similar way as 
\begin{equation}\label{Eq_usVector}
\mathbf{u}_\text{s}(l)\triangleq[\begin{matrix}u_{\text{s},1}(l)&\cdots&u_{\text{s},v}(l)&\cdots&u_{\text{s},N_\text{s}}(l)\end{matrix}]^\text{T}, 
\end{equation}
\noindent which is a look-up table from which the laser power is selected. To define the look-up rule of the laser power, the temporal input command is stacked in a vector 
\begin{equation}\label{Eq_utVector}
\overline{\mathbf{u}}(l)\triangleq[\begin{matrix}u_\text{t}(0, l)&u_\text{t}(1, l)&\cdots&u_\text{t}(N_\text{t}-1, l)\end{matrix}]^\text{T}, 
\end{equation}
\noindent where the controller selects power values $u_\text{t}$ from the spatial look-up table $\mathbf{u}_\text{s}$ by the control law 
\begin{equation}\label{Eq_us2ut}
\overline{\mathbf{u}}(l)=\mathbf{P}(l)\mathbf{u}_\text{s}(l), 
\end{equation}
\noindent where $\mathbf{P}$ can be defined in two ways: the one-step-backward look-up table 
\begin{equation}\label{Eq_P}
\mathbf{P}=\{p_{ij}\}, 
\end{equation}
\begin{equation}\label{Eq_pij}
p_{ij}=\begin{cases}
1 & \text{if }(i-1)\in\mathcal{S}_j\\
0 & \text{otherwise}
\end{cases}
\end{equation}
or the one-step-forward look-up table 
\begin{equation}\label{Eq_pij2}
p_{ij}=\begin{cases}
1 & \text{if }i\in\mathcal{S}_j\\
0 & \text{otherwise}
\end{cases}. 
\end{equation}

\textbf{Example 5}: Consider the example given in Example 4. The one-step-backward $\mathbf{P}$ is 
\begin{equation}\label{Eq_ExP}
\mathbf{P}=
\begin{bmatrix}
1 & 1 & 0 & 0 & 0 & 1 & 1 & 0 & 0 & 0 \\
0 & 0 & 1 & 1 & 1 & 0 & 0 & 0 & 0 & 0 \\
0 & 0 & 0 & 0 & 0 & 0 & 0 & 1 & 1 & 0 \\
0 & 0 & 0 & 0 & 0 & 0 & 0 & 0 & 0 & 1
\end{bmatrix}^\text{T} 
\end{equation}
and the one-step-forward $\mathbf{P}$ is 
\begin{equation}\label{Eq_ExP2}
\mathbf{P}=
\begin{bmatrix}
1 & 0 & 0 & 0 & 0 & 1 & 1 & 0 & 0 & 0 \\
0 & 1 & 1 & 1 & 0 & 0 & 0 & 0 & 0 & 0 \\
0 & 0 & 0 & 0 & 0 & 0 & 1 & 1 & 0 & 0 \\
0 & 0 & 0 & 0 & 0 & 0 & 0 & 0 & 1 & 1
\end{bmatrix}^\text{T}. 
\end{equation}

With the definitions of the spatial and temporal input and output vectors above, by deriving the system propagation directly from (\ref{Eq_system}),(\ref{Eq_ut2u}), the temperature state is 
\begin{equation}\label{Eq_x02xn}
\mathbf{x}(n, l)=\mathbf{A}^n(l)\mathbf{x}(0, l)+\sum_{j=0}^{n-1}\mathbf{A}^{n-j-1}(l)\mathbf{B}(l)\mathbf{h}(j,l)u_\text{t}(j,l). 
\end{equation}
\noindent At the end of each iteration, there is a finalization process for powder recoating where the laser is off. The initial condition of a new iteration is \cite{Wang}
\begin{equation}\label{Eq_recoat}
\mathbf{x}(0, l+1)=\mathbf{\Delta}_1\mathbf{A}_\text{r}(l)\mathbf{x}(N_\text{t}, l), 
\end{equation}
\noindent where $\mathbf{A}_\text{r}$ is the temperature reset operator, which represents the cooling down process with zero inputs, and 
\begin{equation}\label{Eq_reset_x}
\mathbf{A}_\text{r}=\mathbf{0}
\end{equation}
according to the Assumption 2, and $\mathbf{\Delta}_1$ is the initialization of the new layer, namely, 
\begin{equation}\label{Eq_Delta1}
\mathbf{\Delta}_1=\left[
\begin{array}{c|c}
\mathbf{0} & \mathbf{I} \\
\hline
\mathbf{0} & \mathbf{0}
\end{array}
\right], 
\end{equation}
\noindent which has rows of zeros for the new top layer and is an operator that shifts the old layers downwards and initializes the new top layer to 0. Therefore, the layer-to-layer propagation is 
\begin{equation}\label{Eq_xl2xl+1}
\begin{split}
\mathbf{x}(0, l+1)=\mathbf{\Delta}_1\mathbf{A}_\text{r}(l)[\mathbf{A}^{N_\text{t}}(l)\mathbf{x}(0, l)+\\
\sum_{j=0}^{N_\text{t}-1}\mathbf{A}^{N_\text{t}-j-1}(l)\mathbf{B}(l)\mathbf{h}(j,l)u_\text{t}(j,l)]. 
\end{split}
\end{equation}
\noindent To simplify (\ref{Eq_xl2xl+1}), let 
\begin{equation}\label{Eq_AL}
\mathbf{A}_\text{L}(l)=\mathbf{\Delta}_1\mathbf{A}_\text{r}(l)\mathbf{A}^{N_\text{t}}(l), 
\end{equation}
\begin{equation}\label{Eq_BL}
\mathbf{B}_\text{L}(l)=\mathbf{\Delta}_1\mathbf{A}_\text{r}(l)\mathbf{\tilde{B}}(l), 
\end{equation}
\noindent where 
\begin{equation}\label{Eq_Btilde}
\begin{split}
\mathbf{\tilde{B}}(l)=[
\begin{array}{c|c}
\mathbf{A}^{N_\text{t}-1}(l)\mathbf{B}(l)\mathbf{h}(0, l) & \mathbf{A}^{N_\text{t}-2}(l)\mathbf{B}(l)\mathbf{h}(1, l)
\end{array}
\\
\begin{array}{c|c}
\cdots & \mathbf{B}(l)\mathbf{h}(N_\text{t}-1, l)
\end{array}
]. 
\end{split}
\end{equation}
\noindent Therefore, the layer-domain system state equation is 
\begin{equation}\label{Eq_sysLayer}
\mathbf{x}(0, l+1)=\mathbf{A}_\text{L}(l)\mathbf{x}(0, l)+\mathbf{B}_\text{L}(l)\overline{\mathbf{u}}(l). 
\end{equation}
In PBF, the object of control is not the temperature field, but the melt pool measurement. From the measurement equation (\ref{Eq_MeasureLin}), the output is 
\begin{equation}\label{Eq_ytInLayer}
\begin{split}
y_\text{t}(n,l)=\mathbf{f}^\text{T}(n, l)[\mathbf{A}^n(l)\mathbf{x}(0, l)+\\
\sum_{j=0}^{n-1}\mathbf{A}^{n-j-1}(l)\mathbf{B}(l)\mathbf{h}(j,l)u_\text{t}(j,l)]. 
\end{split}
\end{equation}
\noindent Let 
\begin{equation}\label{Eq_CL}
\mathbf{C}_\text{L}(l)=\begin{bmatrix}
\mathbf{f}^\text{T}(1, l)\mathbf{A}(l)\\
\mathbf{f}^\text{T}(2, l)\mathbf{A}^2(l)\\
\vdots\\
\mathbf{f}^\text{T}(N_\text{t}, l)\mathbf{A}^{N_\text{t}}(l)
\end{bmatrix}, 
\end{equation}
\begin{equation}\label{Eq_DL}
\mathbf{D}_\text{L}(l)=\{d_{\text{L},ij}(l)\},
\end{equation}
\noindent where 
\begin{equation}\label{Eq_dLij}
d_{\text{L},ij}(l)=
\begin{cases}
0 & \text{if } i<j
\\
\mathbf{f}^\text{T}(i,l)\mathbf{A}^{i-j}(l)\mathbf{B}(l)\mathbf{h}(j-1,l) & \text{if } i\geq j
\end{cases}.
\end{equation}
\noindent Then the layer-domain output equation is 
\begin{equation}\label{Eq_ytVectorInLayer}
\overline{\mathbf{y}}(l)=\mathbf{C}_\text{L}(l)\mathbf{x}(0,l)+\mathbf{D}_\text{L}(l)\overline{\mathbf{u}}(l). 
\end{equation}

In summary, a layer-domain PBF system model with temporal inputs, $\overline{\mathbf{u}}(l)$, and temporal outputs, $\overline{\mathbf{y}}(l)$, is written compactly as 

\noindent(\textbf{Temporal Layer-Domain System})
\begin{equation}\label{Eq_SyeyemCompInLayer}
\begin{cases}
\mathbf{x}(0, l+1)=\mathbf{A}_\text{L}(l)\mathbf{x}(0, l)+\mathbf{B}_\text{L}(l)\overline{\mathbf{u}}(l)
\\
\overline{\mathbf{y}}(l)=\mathbf{C}_\text{L}(l)\mathbf{x}(0,l)+\mathbf{D}_\text{L}(l)\overline{\mathbf{u}}(l).
\end{cases}
\end{equation}
A layer-domain PBF model with spatially-defined ``look-up table" inputs, (\ref{Eq_us2ut}), and voxel feature map outputs, $\mathbf{y}_\text{s}(l)$, is written compactly as

\noindent(\textbf{Spatial Layer-Domain System})
\begin{equation}\label{Eq_SyeyemCompInLayerSp}
\begin{cases}
\mathbf{x}(0, l+1)=\mathbf{A}_\text{L}(l)\mathbf{x}(0, l)+\mathbf{B}_\text{L}(l)\mathbf{P}(l)\mathbf{u}_\text{s}(l)
\\
\mathbf{y}_\text{s}(l)=\mathbf{Q}(l)\mathbf{C}_\text{L}(l)\mathbf{x}(0,l)+\mathbf{Q}(l)\mathbf{D}_\text{L}(l)\mathbf{P}(l)\mathbf{u}_\text{s}(l).
\end{cases}
\end{equation}
\noindent Combining the temperature-resetting condition (\ref{Eq_reset_x}) between layers with the forward-calculating the temporal state dynamics, the temporal layer-domain system is reduced to

\noindent(\textbf{Temporal Input-Output Layer-Domain System})
\begin{equation}\label{Eq_ut2yt}
\overline{\mathbf{y}}(l)=\mathbf{D}_\text{L}(l)\overline{\mathbf{u}}(l). 
\end{equation}
\noindent The spatial layer-domain system in (\ref{Eq_SyeyemCompInLayerSp}) is reduced to 

\noindent(\textbf{Spatial Input-Output Layer-Domain System})
\begin{equation}\label{Eq_us2ys}
\mathbf{y}_\text{s}(l)=\mathbf{G}_\text{s}(l)\mathbf{u}_\text{s}(l), 
\end{equation}
\noindent where 
\begin{equation}\label{Eq_G_s}
\mathbf{G}_\text{s}(l)=\mathbf{Q}(l)\mathbf{D}_\text{L}(l)\mathbf{P}(l). 
\end{equation}
\textbf{Remark}: Equation (\ref{Eq_us2ys}) yields a new layer-domain, purely spatial form of the spatial-temporal heat transfer model, in which the inputs and outputs are defined directly in a voxel map. Path and feature measurement definitions, along with all temporal dynamics, are embedded in $\mathbf{G}_\text{s}$, which is reduced to the pure spatial relationship between voxel inputs and voxel outputs. Notably, the dimensions of $\mathbf{y}_\text{s}$, $\mathbf{u}_\text{s}$, $\mathbf{G}_\text{s}$ and the voxel grid are independent of the spatial-temporal finite difference grid and, thus, a reduced resolution voxel grid does not constitute any necessary loss in the fidelity of the numerical calculation of the underlying physics. Further, (\ref{Eq_us2ys}) provides a different perspective of PBF control models and a convenient form suitable for spatial layer control. 

\section{Layer-Domain Output Controllability}
To analyze controllability in the layer domain, the system in (\ref{Eq_SyeyemCompInLayer}) can be treated as a Linear Time-Variant (LTV) system where the layer, $l$, plays the role of ``time.'' The solution is \cite{Ludyk, Wieberg} 
\begin{equation}\label{Eq_LTV_solution}
\mathbf{x}(0,l)=\mathbf{\Phi}(l,0)\mathbf{x}(0,0)+\sum_{r=0}^{l-1}\mathbf{\Phi}(l, r+1)\mathbf{B}_\text{L}(r)\overline{\mathbf{u}}(r), 
\end{equation}
\begin{equation}\label{Eq_LTV_solution_output}
\begin{gathered}
\overline{\mathbf{y}}_\text{s}(l)=\mathbf{C}_\text{L}(l)\mathbf{\Phi}(l,0)\mathbf{x}(0,0)\\
+\sum_{r=0}^{l-1}\mathbf{C}_\text{L}(l)\mathbf{\Phi}(l, r+1)\mathbf{B}_\text{L}(r)\overline{\mathbf{u}}(r)+\mathbf{D}_\text{L}(l)\overline{\mathbf{u}}(l), 
\end{gathered}
\end{equation}
where $\mathbf{\Phi}$ is the well-known state-transition matrix 
\begin{equation}\label{Eq_state_transition}
\mathbf{\Phi}(l_1,l_2) =\begin{cases}
\prod_{l=l_2}^{l_1-1}\mathbf{A}_\text{L}(l) & \text{if }l_1>l_2\\
\mathbf{I} & \text{if }l_1=l_2
\end{cases}. 
\end{equation}

 A sufficient condition for output controllability of (\ref{Eq_SyeyemCompInLayer}) is introduced in the following theorem: 

\vspace{6pt}
\noindent\textbf{Theorem 1 (Temporal Layer-Domain System Output Controllability):} For the temporal layer-domain system, (\ref{Eq_SyeyemCompInLayer}), the system is output-controllable if $\mathbf{D}_\text{L}$ is full-rank, namely, 
\begin{equation}
\text{det}(\mathbf{D}_\text{L}(l))=\prod_j\mathbf{f}^\text{T}(j,l)\mathbf{B}(l)\mathbf{h}(j-1, l)\neq 0, 
\end{equation}
for every layer $l$. 
\vspace{6pt}

\noindent\textbf{Proof:} Consider the Linear Time-Invariant (LTI) case of (\ref{Eq_LTV_solution}),(\ref{Eq_LTV_solution_output}) in which $\mathbf{A}_\text{L}, \mathbf{B}_\text{L}, \mathbf{C}_\text{L}$ and $\mathbf{D}_\text{L}$ are layer-invariant. Then, the LTI  case is output controllable over the layer interval $[0, l_1]$ if the output controllability matrix 
\begin{equation}\label{Eq_MOC}
\begin{gathered}
\mathbf{M}_\text{OC}=[
\begin{array}{c|c|}
\mathbf{C}_\text{L}(l_1)\mathbf{B}_\text{L}(l_1) &
\mathbf{C}_\text{L}(l_1)\mathbf{\Phi}(l_1,l_1-1)\mathbf{B}_\text{L}(l_1)
\end{array}\\
\begin{array}{c|c|c}
\cdots&
\mathbf{C}_\text{L}(l_1)\mathbf{\Phi}(l_1,1)\mathbf{B}_\text{L}(l_1) & \mathbf{D}_\text{L}(l_1)
\end{array}
]
\end{gathered}
\end{equation}
\noindent is full rank. Now, because $\mathbf{D}_\text{L}$ is square, the condition $\text{det}(\mathbf{D}_\text{L})\neq 0$ is equivalent to $\mathbf{D}_\text{L}$ having full rank. It follows directly that the output controllability matrix $\mathbf{M}_\text{OC}$ is full row rank for any $l_1>0$, and the LTI case is output controllable. Finally, because output-controllability is achieved even for $l_1=1$, the LTV system is always output controllable over one layer-step, at every layer and, therefore, is also output controllable. \hfill $\blacksquare$
\vspace{6pt}

\textbf{Remark}: In a PBF system, $\mathbf{f}^\text{T}(j, l)\mathbf{B}(l)\mathbf{h}(j-1, l)$ is the gain from the input indexed by time $j-1$ to the output indexed by time $j$ in layer $l$. The sufficient condition is usually satisfied easily since the melt pool measurement is usually non-zero when the laser power is non-zero. 

As discussed by the previous remark, output controllability of the temporal system is generally assured. However, as demonstrated by the following theorems, output controllability of the transformed problem into the spatial voxel domain, in which the output is the voxel feature map of interest to the manufacturing community, does not automatically follow. 

\vspace{6pt}
\noindent\textbf{Theorem 2 (Spatial Input-Output Layer-Domain Output Controllability):} 
Consider the spatial input-output layer domain system in (\ref{Eq_us2ys}),(\ref{Eq_G_s}). The system is output controllable if $\mathbf{G}_\text{s}(l)$ has full row-rank for all $l$. 

\vspace{6pt}
\noindent The proof follows similarly to the Theorem 1 above. 

\vspace{6pt}
\noindent\textbf{Remark:} The matrix $\mathbf{Q}$ having full row-rank is a necessary condition for $\mathbf{G}_\text{s}$ to have full row-rank. This is trivially satisfied when each voxel has a measurement assigned to it, that is, when $\mathcal{S}_i\neq\emptyset$ for all $i$. Also, $\mathbf{P}$ always has full column-rank because $\mathbf{P}$ defines the input path and there is always an input in each voxel included in the input path. However, $\mathbf{Q}$ and $\mathbf{P}$ having full rank are necessary but not sufficient for $\mathbf{G_s}$ to have full rank. 

Although the temporal output controllability is usually satisfied naturally by the physics of the PBF process, the spatial output controllability will depend, in general, on the choice of $\mathbf{P}$ and $\mathbf{Q}$. A more direct condition will be established for a common choice of $\mathbf{P}$ and $\mathbf{Q}$. Before presenting that theorem, the following definition, lemma, and corollary are necessary.

\vspace{6pt}
\noindent\textbf{Definition:} A square matrix $\mathbf{M}=\{m_{ij}\}$ is said to be strictly diagonally dominant if 
\begin{equation}
|m_{ii}|>\sum_{j\neq i}|m_{ij}|
\end{equation}
for all $i$. 

\vspace{6pt}
\noindent\textbf{Lemma:} Every strictly diagonally dominant matrix is invertible. 

\vspace{6pt}
\noindent\textbf{Proof:}
For any strictly diagonally dominant $\mathbf{M}$, suppose that it is not invertible. Then there exists a vector $\mathbf{w}\neq \mathbf{0}$ such that 
\begin{equation}
\mathbf{Mw}=\mathbf{0}. 
\end{equation}
Let $w_i$ be the element of $\mathbf{w}$ with the greatest absolute value. Then 
\begin{equation}
\sum_{j}m_{ij}w_j=0, 
\end{equation}
\begin{equation}
m_{ii}=-\sum_{j\neq i}\frac{m_{ij}w_j}{w_i}, 
\end{equation}
\begin{equation}
|m_{ii}|\leq\sum_{j\neq i}|m_{ij}\frac{w_j}{w_i}|\leq|\sum_{j\neq i}|m_{ij}|, 
\end{equation}
which is a contradiction, so every strictly diagonally dominant matrix is invertible. \hfill $\blacksquare$

\vspace{6pt}
\noindent\textbf{Corollary:} 
The spatial input-output layer domain system in (\ref{Eq_us2ys}),(\ref{Eq_G_s}) is output controllable if $\mathbf{G}_\text{s}(l)$ is strictly diagonally dominant for all $l$. 

\vspace{6pt}
\noindent\textbf{Remark:} Although diagonal dominance is a much stronger condition than full rank, $\mathbf{G}_\text{s}$ satisfies it easily if the material cooling rate is relatively fast, the sampling rate is relatively slow and the voxel spacing is relatively large. A theorem is now presented regarding the material cooling rate being ``relatively fast". 

\vspace{6pt}
\noindent\textbf{Theorem 3 (Fast-Cooling Output Controllability):} With the one-step-forward $\mathbf{P}$ in (\ref{Eq_pij2}) and the average-within-the-voxel operator $\mathbf{Q}$ in (\ref{Eq_Q}),(\ref{Eq_qij}), the spatial input-output layer domain system in (\ref{Eq_us2ys}),(\ref{Eq_G_s}) is output-controllable if $\mathbf{D}_\text{L}$ has no negative entries and is strictly diagonally dominant. 

\vspace{6pt}
\noindent\textbf{Proof:} 
With the one-step-forward $\mathbf{P}$ and the average-within-the-voxel operator $\mathbf{Q}$, it is straight-forward to find 
\begin{equation}
\mathbf{Q}=\mathbf{RP}^\text{T}, 
\end{equation}
where
\begin{equation}
\mathbf{R} = 
\left[
\begin{matrix}
|\mathcal{S}_1| & & &\\
& |\mathcal{S}_2| & & \\
& & \ddots &\\
& & & & |\mathcal{S}_{N_\text{s}}|
\end{matrix}\right]. 
\end{equation}
Then 
\begin{equation}
\mathbf{G}_\text{s}=\mathbf{RP}^\text{T}\mathbf{D}_\text{L}\mathbf{P}. 
\end{equation}
Because $\mathbf{R}$ is diagonal and, thus, invertible, $\mathbf{G}_\text{s}$ is full rank if and only if $\mathbf{P}^\text{T}\mathbf{D}_\text{L}\mathbf{P}$ is invertible. 

Let $\mathbf{M}=\{m_{ij}\}=\mathbf{P}^\text{T}\mathbf{D}_\text{L}\mathbf{P}$. With the specific structure of the one-step-forward $\mathbf{P}$, 
\begin{equation}
m_{ij}=\sum_{j'\in\mathcal{S}_{j}}\sum_{i'\in\mathcal{S}_{i}}d_{\text{L},i'j'}. 
\end{equation}
Due to the construction of $\mathbf{P}$, $\mathcal{S}_i\cap\mathcal{S}_j=\emptyset$ for $i\neq j$. Therefore, when $\mathbf{D}_\text{L}$ has no negative entries and is strictly diagonally dominant, 
\begin{equation}
|m_{ii}|=\sum_{j'\in\mathcal{S}_{i}}\sum_{i'\in\mathcal{S}_{i}}d_{\text{L},i'j'}>\sum_{j'\notin\mathcal{S}_{i}}\sum_{i'\in\mathcal{S}_{i}}d_{\text{L},i'j'}=\sum_{j\neq i}|m_{ij}|, 
\end{equation}
so $\mathbf{M}$ is also strictly diagonally dominant and, thus, $\mathbf{G}_\text{s}$ is also strictly diagonally dominant. Therefore, the spatial input-output layer-domain system is output controllable. \hfill $\blacksquare$

\vspace{6pt}
\noindent\textbf{Remark:} 
According to its structure (\ref{Eq_DL}), the entry $d_{\text{L}, ij}$ is the temperature response having decayed after $(i-j)$ time steps. Therefore, in general, the dominance of the diagonal entry depends on the decay rate in time and space between the sample points $i$ and $j$, although geometry is a third factor. Especially, if it is true for all $i, j$ that 
\begin{equation}
\frac{d_{\text{L}, i(j-1)}}{d_{\text{L}, ij}}<\frac{1}{2}, 
\label{Eq_D_remark}
\end{equation}
then
\begin{equation}
\begin{split}
d_{\text{L}, i(j-1)}+d_{\text{L}, i(j-2)}+d_{\text{L}, i(j-3)}+\cdots \\
<d_{\text{L}, ij}(\frac{1}{2}+\frac{1}{4}+\frac{1}{8}+\cdots)<d_{\text{L}, ij},   
\end{split}
\end{equation}
and, thus, $\mathbf{D}_\text{L}$ is strictly diagonally dominant. 

\begin{figure}[htbp]
\centering 
\includegraphics[width = 3in]{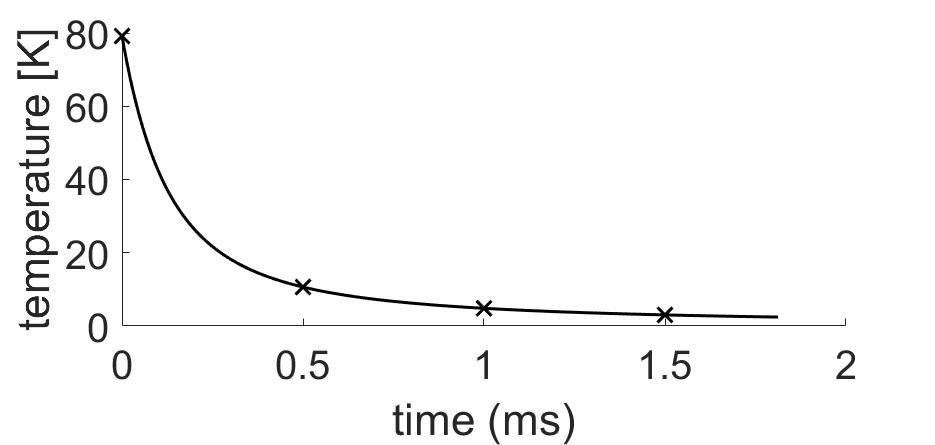}
\caption{Temperature Response to Laser Pulse at Right Angle Corner}
\label{Fig_fast_cool}
\end{figure}

In the experiments conducted in this paper, the diagonal dominance of $\mathbf{D}_\text{L}$ is coarsely evaluated, although $\mathbf{D}_\text{L}$ is not identified. The sampling rate is 2 kHz, the sample distance is 400 μm and the hatch spacing is 100 μm. The conductivity and diffusivity of solid steel is 33.5 W$/$(mK) and $6\times 10^{-6} \text{m}^2/\text{s}$, respectively. Consider the point at the corner of the right angle of a steel part, with adiabatic conditions on the boundaries. In this case, when a 1 W laser pulse strikes the corner, the temperature at the corner cools down as shown in Fig.~\ref{Fig_fast_cool}. With the relatively slow sampling rate of 2 kHz, the diagonal dominance of $\mathbf{D}_\text{L}$ is assured. While generally a fast sampling rate is often desired in control, here a relatively slow sampling rate of the camera provides a convenient assurance of controllability, because a relatively slow sampling rate makes the adjacent two samples differs enough to satisfy (\ref{Eq_D_remark}). 

\section{Experimental Results}
Spatial Iterative Learning Control (SILC), which regulates outputs in a 2D plane by learning from previous inputs in a repetitive process, has been applied to additive manufacturing processes \cite{Hoelzle4, Hoelzle6, Hoelzle10}. With the spatial layer-to-layer PBF model above, SILC can be implemented to control spatial voxel feature maps of melt pool features to achieve a uniform output. In PBF, a uniform input power often does not produce uniform melt pools over the layer space due to the non-uniformity of the part geometry, especially at path turnaround points, corners and overhang edges. Experiments are conducted to demonstrate the capability of SILC to reduce the non-uniformity of melt pool features in PBF. 

The experiments in this work are conducted on the open-architecture PBF machine in Rensselaer Polytechnic Institute (RPI). The PBF machine is equipped with a 400W NdYAG laser and a SCANLAB intelliSCANde20 scanner and can build parts with cross-sections as large as $50\times50$ mm$^2$. The machine uses commercially available powder such as stainless steel. For monitoring, a coaxial optical camera (basler acA2000-165μm) with a near-infrared filter (800-950 nm) is used to capture 8-bit intensity images of the melt pool during the process. All images are acquired at a rate of 2kHz, with a size of $64\times64$ pixels and 20 μm per pixel. 
\begin{figure}[htbp]
\centering 
\includegraphics[width = 3in]{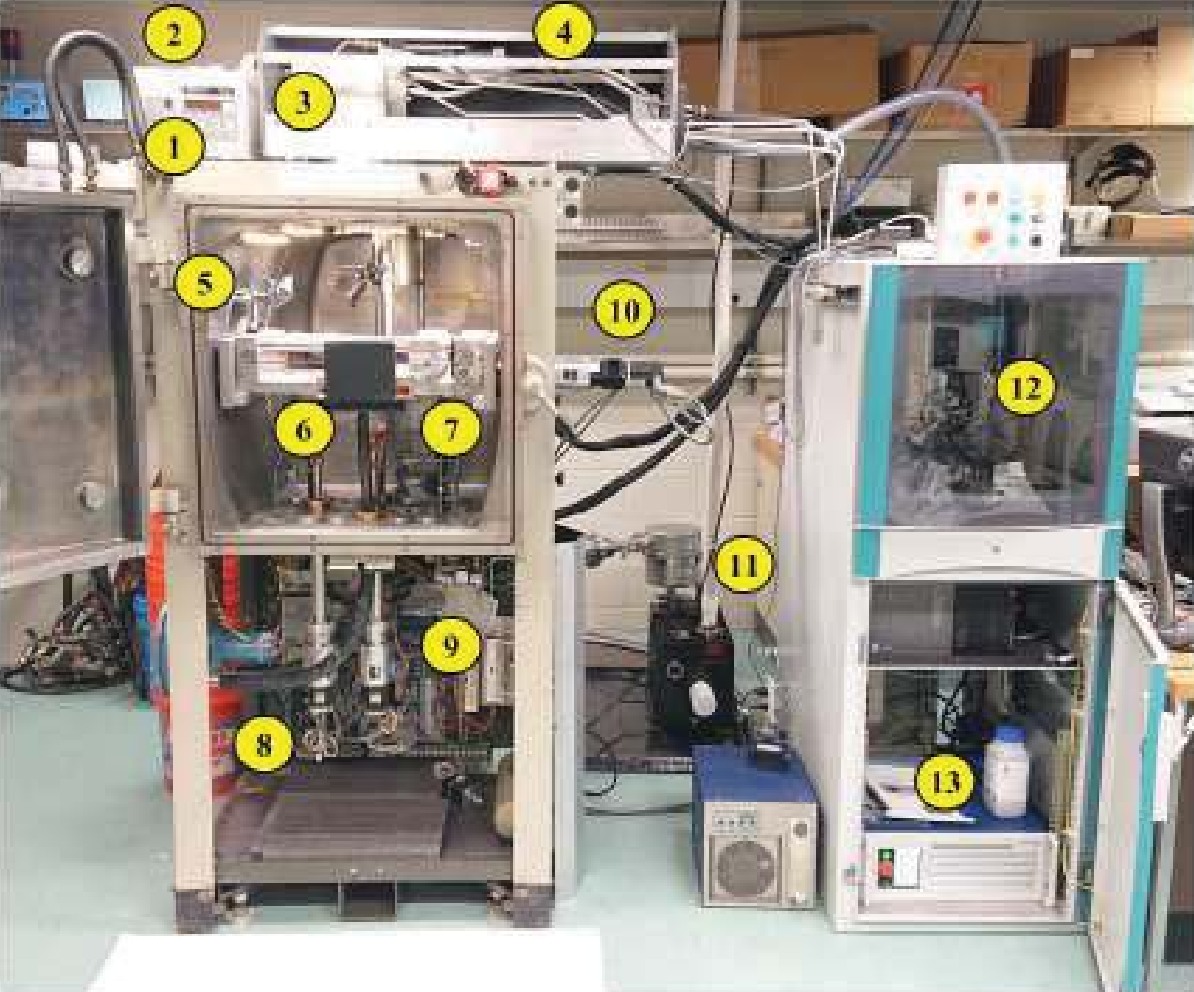}
\caption{Open-Source SLM Machine (1 oxygen analyser, 2 off-axis camera (not visible), 3 scanner, 4 coaxial camera, 5 mirrors for the off-axis camera, 6 dust collector, 7 recoater, 8 powder piston (left) and
build platform piston (right), 9 motion controller, 10 recoater drive, 11 roughing pump, 12 industrial PC cabinet, 13 laser)}
\label{Fig_RPI_machine}
\end{figure}

The control of the PBF system is implemented at two levels. High level supervisory controls are done through AmericaMakes software \cite{America Makes}, augmented with custom C++ codes to provide additional functionality, such as the SILC algorithm. Low-level controls, e.g., laser firing or path execution, are implemented through the SCANLAB RTC 5 scanner control board. Prior to the scanning the process, the scanning instructions for a given layer are formatted as an XML file of scan lines in terms of scan coordinates, scan velocities and power values. Although scan paths and velocities are pre-determined, the supervisory control is capable of overwriting power values at each time step (image acquisition period, i.e., 0.5 ms). For this study, power values are updated on a layer-to-layer basis, where the camera measurements for an entire layer are used to compute the power profile for the subsequent layer. 

\subsection{Spatial ILC - Constant Layer Geometry and Path}
\begin{figure}[hb]
\centering 
\includegraphics[width = 3in]{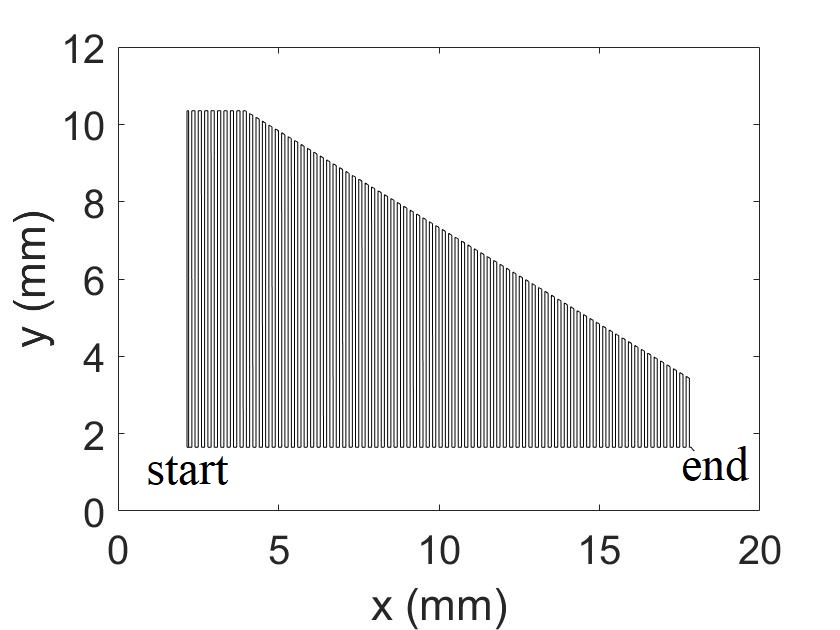}
\caption{Laser Path for Triangular Prism Shape}
\label{Fig_Path1}
\end{figure}

A prism-shape part (Fig.~\ref{Fig_Path1}) is utilized in the first experimental study. In this study, the path for each layer is identical. An 8-bit infrared camera is installed to track melt pools at a frequency of 2 kHz. The laser speed is 0.8 m/s and the hatch spacing is 100 μm. The melt pool area is selected as the output measurement of interest, and is measured by the sum of pixels above a threshold of 100. The spatial voxel feature map is constructed from the output measurements using the average-within-the-voxel method.
\begin{figure*}[ht]
\centerline{\includegraphics[width=7in]{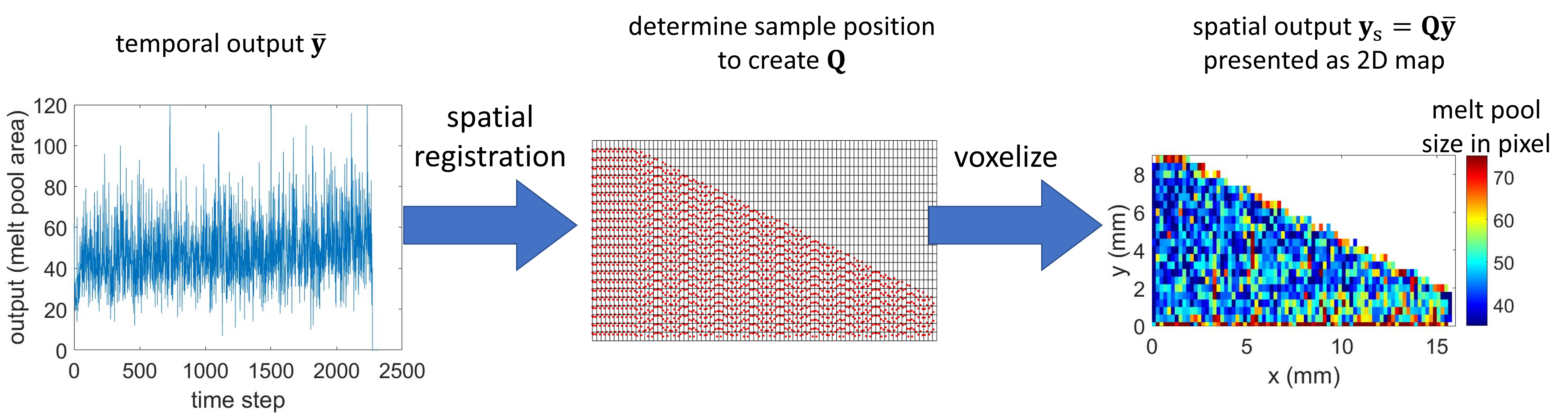}}
\caption{Temporal Signal Voxelized into Voxel Map}
\label{Fig_voxelize}
\end{figure*}
\begin{figure*}[ht]
\centerline{\includegraphics[width=7in]{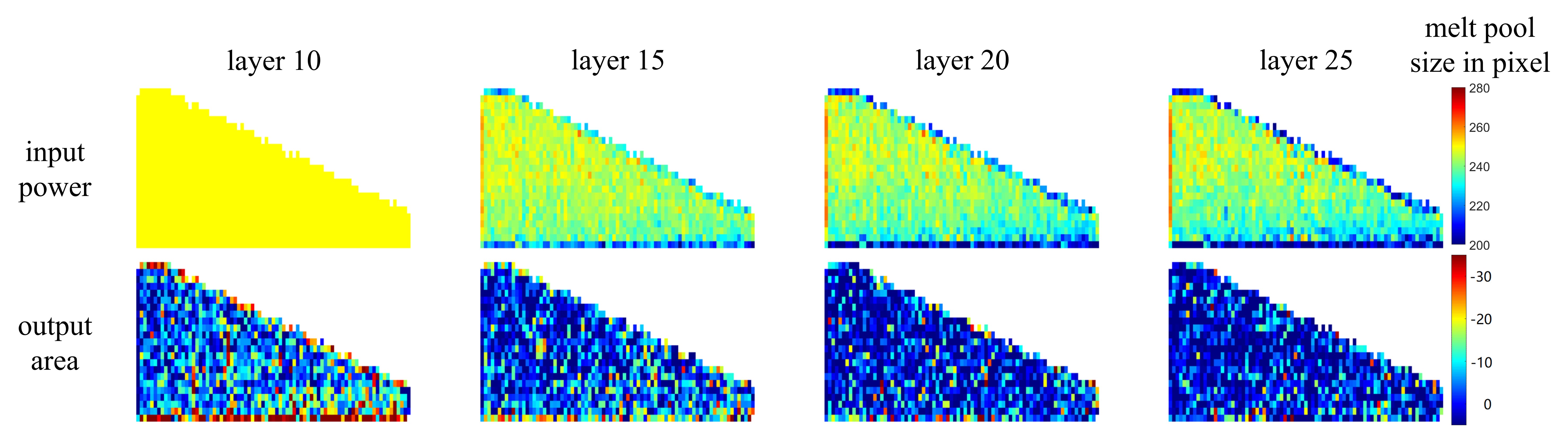}}
\caption{Spatial Control Maps of Inputs and Outputs in Control Experiment of Triangular Prism Part}
\label{Fig_Result}
\end{figure*}

Based on the spatial PBF modeling framework above, a simple SILC law is proposed as
\begin{equation}\label{Eq_ILC1}
\mathbf{u}_\text{s}(l+1)=\mathbf{u}_\text{s}(l)+\gamma \mathbf{e}_\text{s}(l),
\end{equation}
\begin{equation}\label{Eq_ILC2}
\mathbf{e}_\text{s}(l)=\mathbf{y}_\text{d}-\mathbf{y}_\text{s}(l),
\end{equation}
where $\mathbf{u}_\text{s}$ and $\mathbf{y}_\text{s}$ are voxel input and output, respectively, defined above, $\mathbf{e}_\text{s}$ is the voxel output error, $\mathbf{y}_\text{d}$ is the desired reference and the learning gain is tuned to be $\gamma=0.2$. The reference is 40 pixels, which is the melt pool measurement in the central region of the part fabricated with a constant laser power of 250 W. 

The data processing method is illustrated in Fig.~\ref{Fig_voxelize}. The spatial registration determines which sample belongs to which voxel and, accordingly, the average-within-the-voxel operator $\mathbf{Q}$ is created. For a better visualization, the spatial output vector $\mathbf{y}_\text{s}$ is rearranged and plotted in a 2D voxel map; however, it should be noted that it is actually a 1D vector in the control algorithm. 

Results for the experiment are shown in Fig.~\ref{Fig_Result}, with the SILC law applied beginning at layer 10 in order to reduce the effects of the first few layers due to substantial heat transfer to the build plate, powder not being fully compacted, etc. For the first 10 layers P = 250 W. While the output error plot shows significant noise emblematic of optical feedback in PBF, it is notable on layer 10 (before control) that the upper and lower edges show consistently hotter edges as compared to the part interior, and a general increase in heating in the rightmost part section where the paths are relatively short. By applying SILC (\ref{Eq_ILC1}),(\ref{Eq_ILC2}), the input map is adjusted to decrease the laser power on the edges and in the rightmost part section, greatly improving the uniformity of the spatial feature map.

\begin{figure}[htbp]
\centering 
\includegraphics[width = 2.5 in]{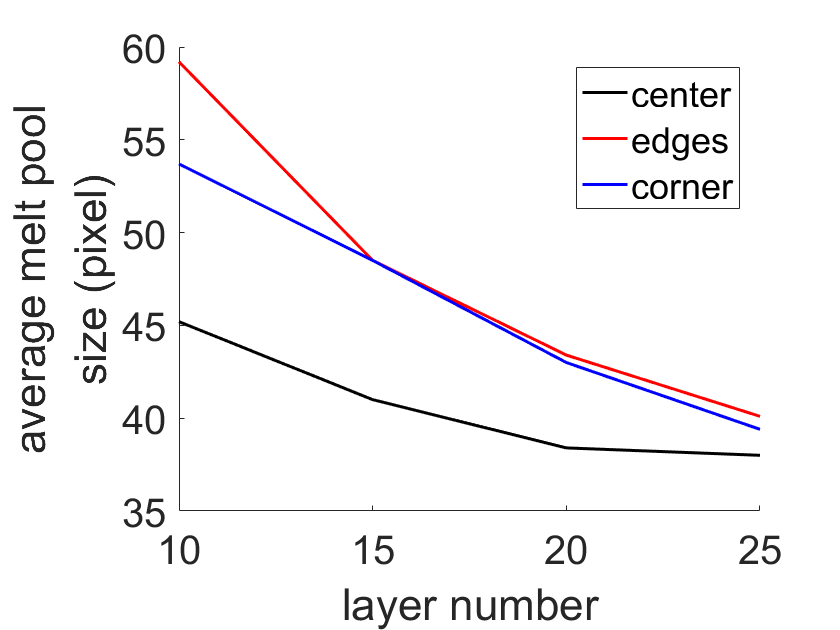}
\caption{Average Melt Pool Sizes (Pixels) in Different Regions}
\label{Fig_regional_avg}
\end{figure}

\begin{figure}[htbp]
\centering 
\includegraphics[width = 2.5 in]{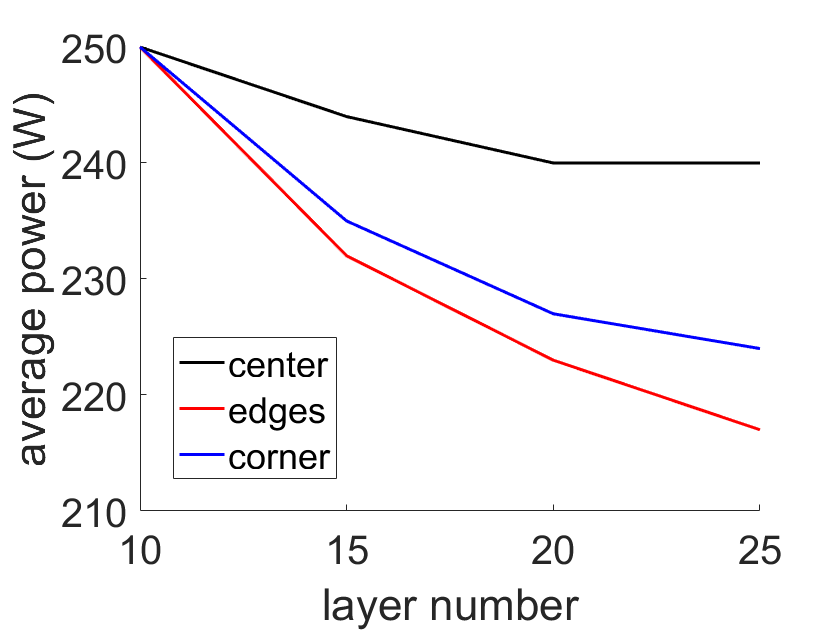}
\caption{Average Laser Power (W) in Different Regions}
\label{Fig_regional_pwr}
\end{figure}

Average melt pool sizes in different regions in Fig.~\ref{Fig_Result} are plotted in Fig.~\ref{Fig_regional_avg}, where the edges refer to the upper and lower edges, the corner refers to the rightmost ten laser tracks and the center refers to the remaining region. The melt pools sizes in the three regions were well above 40 pixels at layer 10. By layer 20, the melt pool sizes in all three regions are within 10\% of the reference. The average laser powers are plotted in Fig.~\ref{Fig_regional_pwr}. All three regions required powers less than 250 W, particularly at the edges and in the corner, to maintain a melt pool size of approximately 40 pixels. 

\subsection{Spatial ILC - Variant Layer Geometry and Path}
\begin{figure}[htbp]
\centering 
\includegraphics[width = 3in]{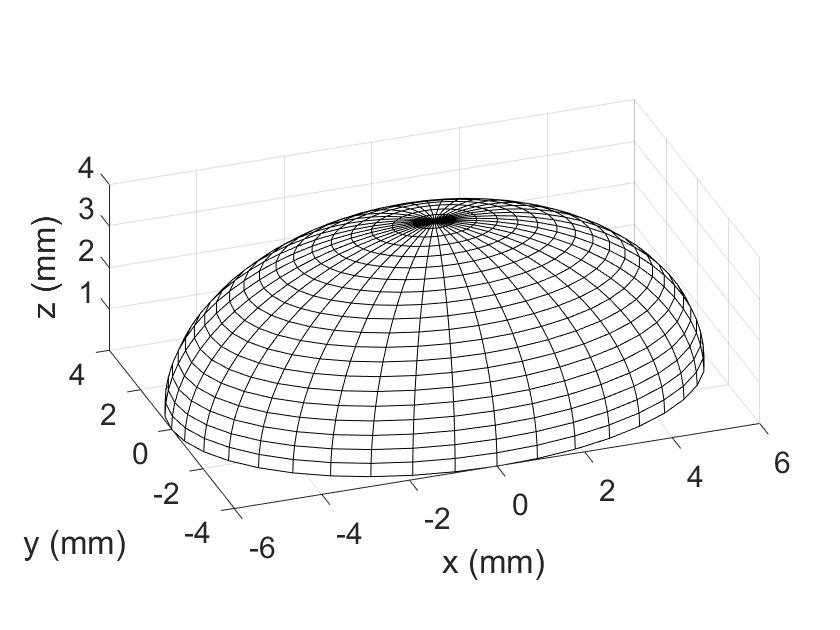}
\caption{Half Ellipsoid Part used in Second Spatial Control Experiment}
\label{Fig_HalfEllipsoid}
\end{figure}

In the second experiment a half ellipsoid 12 mm in length, 8 mm in width and 4 mm in height, as shown in Fig.~\ref{Fig_HalfEllipsoid}, is printed. The laser paths rotate 67° each layer. These layer-wise changes introduce significant variations in the temporal history of measurements and signal lengths, making temporal layer-to-layer correction difficult. The controller in (\ref{Eq_ILC1}),(\ref{Eq_ILC2}) is again implemented. 

An open-loop experiment and a closed-loop experiment are conducted and compared. The nominal laser power is 250 W. After a new configuration of the camera aperture, the reference output is 75, which is the melt pool measurement in the central region of the part when using the nominal power, and the control gain is tuned to be $\gamma=0.25$.

\begin{figure*}
\centerline{\includegraphics[width=7in]{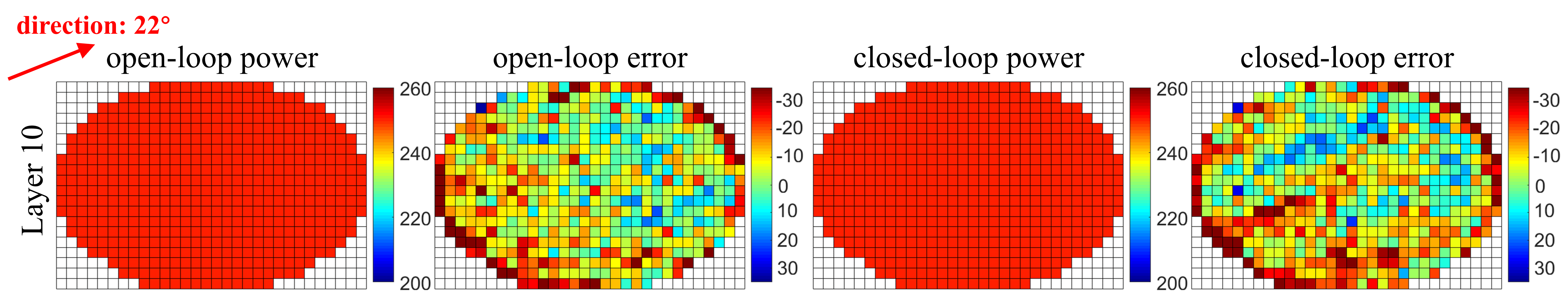}}
\centerline{\includegraphics[width=7in]{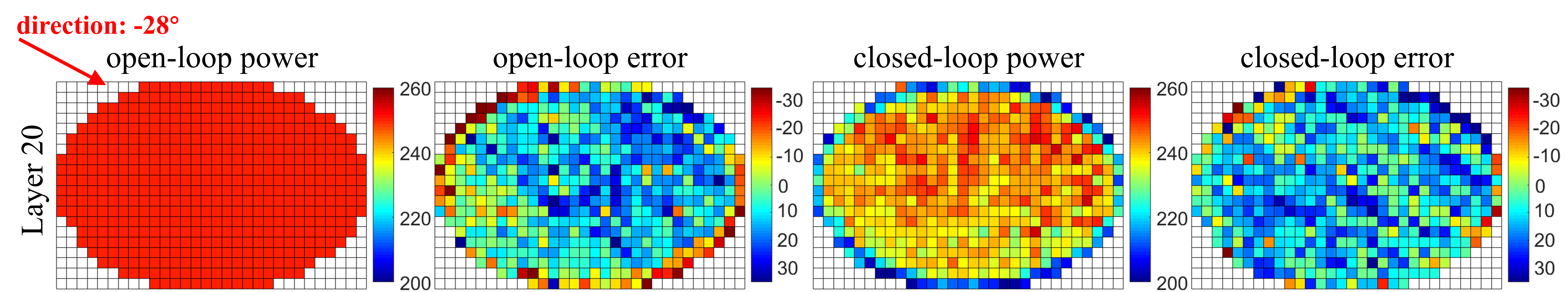}}
\centerline{\includegraphics[width=7in]{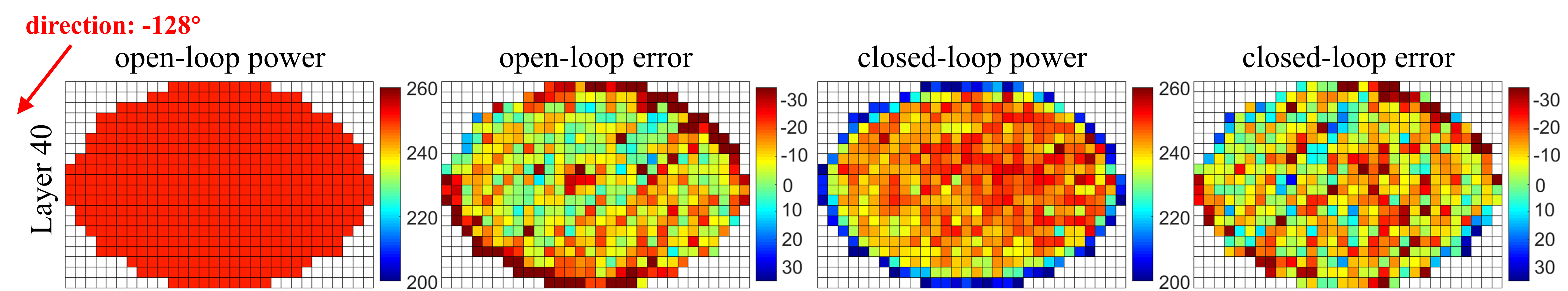}}
\centerline{\includegraphics[width=7in]{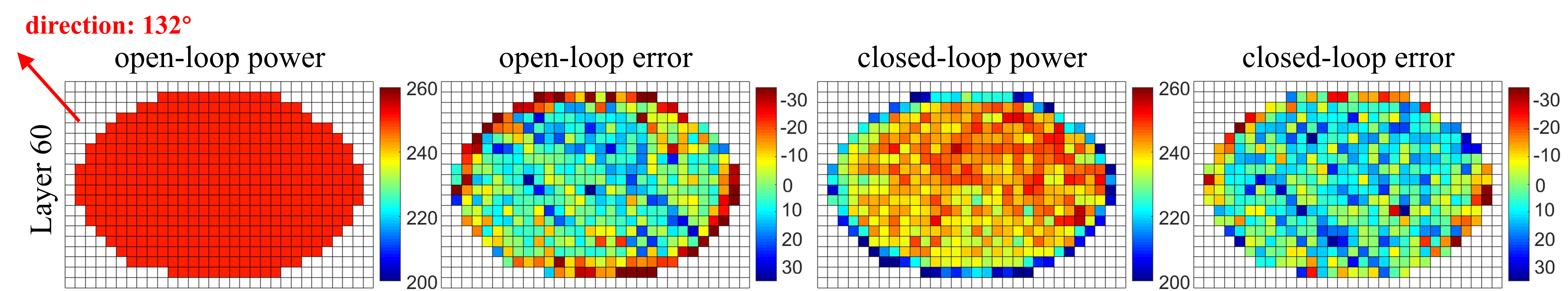}}
\centerline{\includegraphics[width=7in]{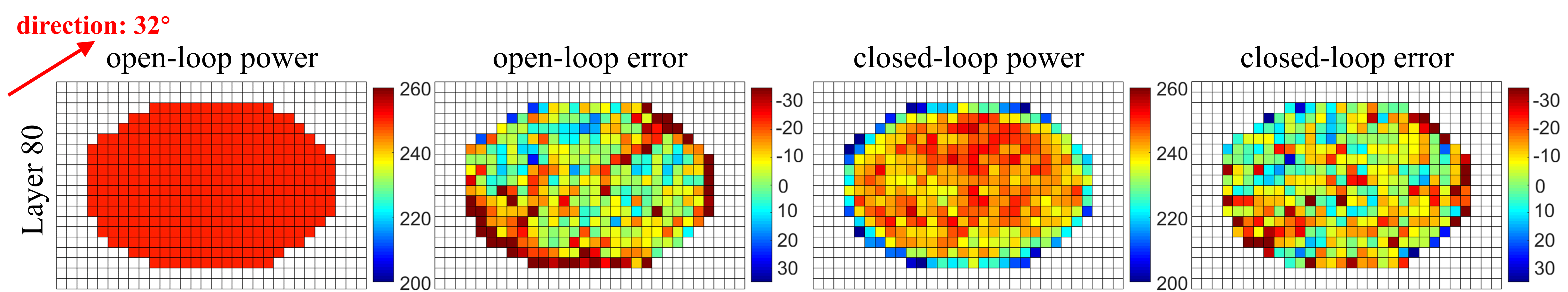}}
\caption{Experimental Results for Half Ellipsoid Part: (From Left to Right) Open-Loop Power and Error and Controlled Power and Error}
\label{Fig_Data10}
\end{figure*}

Experimental results are plotted in Fig.~\ref{Fig_Data10}. In each plot, the path angles with respect to the positive x axis are also labelled. As in the previous example, it is seen that the SILC controller reduces power at the edges of the part, mitigating the hot-edge effect seen in the open-loop response and creating a more uniform spatial feature map. It is notable that such uniformity is achieved despite the changing profile of each layer and the rotation of the raster paths.

\begin{figure*}
\centerline{\includegraphics[width=7in]{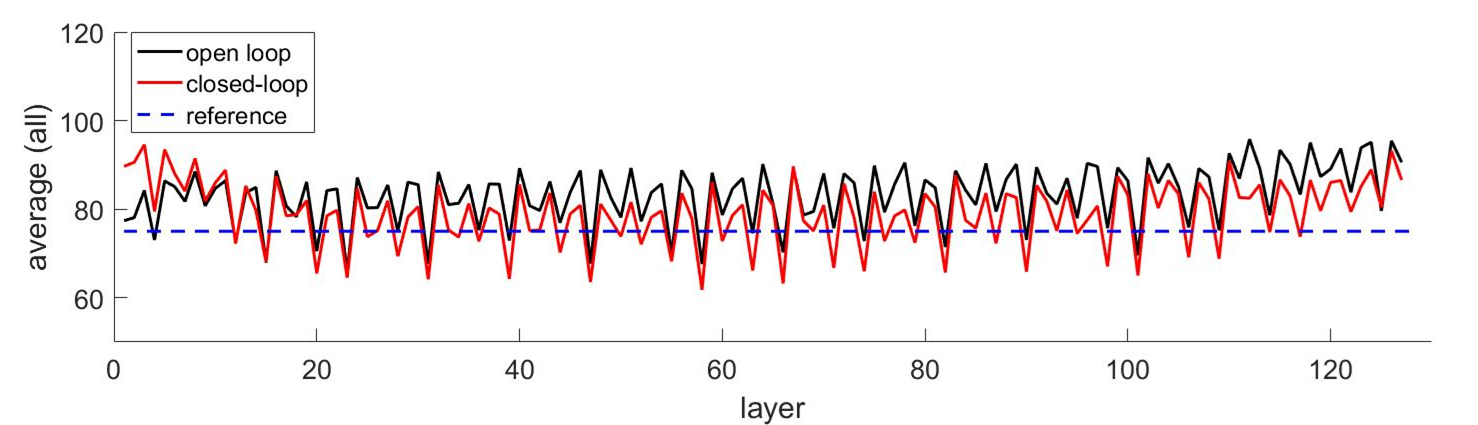}}
\caption{Experimental Results for Half Ellipsoid Part: Average of All Outputs in Each Layer}
\label{Fig_DataAvgAll}
\end{figure*}

\begin{figure*}
\centerline{\includegraphics[width=7in]{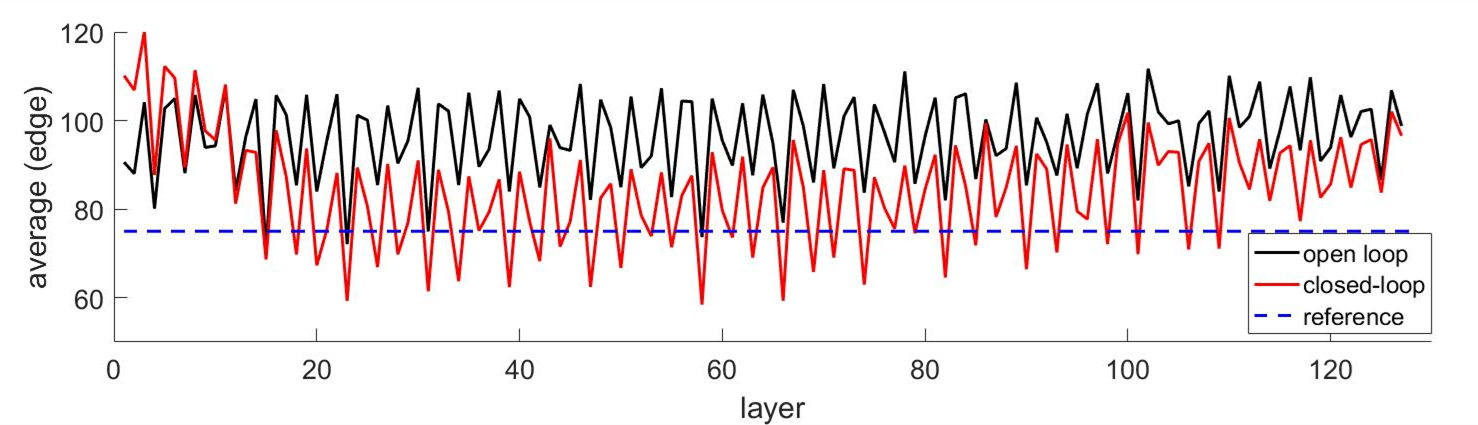}}
\caption{Experimental Results for Half Ellipsoid Part: Average of Edge Outputs in Each Layer}
\label{Fig_DataAvgEdg}
\end{figure*}

The average spatial outputs at each layer are plotted in Fig.~\ref{Fig_DataAvgAll}, and then again in Fig.~\ref{Fig_DataAvgEdg} after isolating only the edge points within each layer. Again, the controller is first implemented after layer 10. The modest improvement of the closed-loop controller on the former is a result of the already tuned open-loop process achieving nearly desirable spatial voxel feature outputs in the center region of the part, the number of which dominate the metric. However, by isolating the edges (Fig.~\ref{Fig_DataAvgEdg}), a dramatic improvement can be seen where the open-loop process performs poorly.

Note in Fig. \ref{Fig_DataAvgAll} and Fig.~\ref{Fig_DataAvgEdg} that the open-loop curve (black) and the closed-loop curve (red) fluctuate consistently with the same ``frequency". Rotating in the oval layers, melt pools get hotter on the longer paths and cooler on the shorter paths. At the beginning layers, the cross-sectional area decreases slowly; however, the cross-sectional area decreases very quickly towards the end of the print. After about 80 layers, the melt pool area increases because the cross-section shrinks fast and the edge of a new layer is not laid on the edge of the previous edge. Then the SILC cannot learn the edge pattern by learning the voxel outputs vertically.


\section{Summary and Conclusions}
In this paper a spatial layer-to-layer control-oriented thermal PBF model was developed. System output controllability was analyzed and sufficient conditions for output controllability were derived. A SILC law was constructed and applied to regulate melt pool geometry in two experiments, a prism where the path and part geometry were constant for each layer and a half ellipsoid where the path and part geometry changed each layer. These results were compared to the corresponding results from experiments using constant process parameters.

The controllability analysis proved that the temporal layer-domain system and the spatial input-output layer-domain system are output controllable if the matrices $\mathbf{D}_\text{L}$ (i.e., the temporal layer-domain system matrix relating the output to the input) and $\mathbf{G}_\text{s}$ (i.e., the spatial layer-domain system matrix relating the output to the input), respectively, are full rank. The experimental results demonstrated that the SILC control law was able to regulate melt pool area at a constant value, even at the part boundaries where excess heat tends to increase melt pool size. Further, the SILC law was able to regulate the melt pool area even when the path and part geometry changed every layer. The layer-to-layer spatial modeling framework was able to capture spatial-driven physics (i.e., heat build up at part boundaries) and provided a basis to construct a controller that could regulate these physics. The framework directly incorporates pathing, sensing and actuation locations and part geometry, making it well-suited for voxel-level monitoring and characterization techniques.



\begin{thebibliography}{00}
\bibitem{Singh}Singh, D. Dev, T. Mahender, and Avala Raji Reddy. ``Powder bed fusion process: A brief review." Materials Today: Proceedings 46 (2021): 350-355.

\bibitem{McCann}McCann, Ronan, Muhannad A. Obeidi, Cian Hughes, Eanna McCarthy, Darragh S. Egan, Rajani K. Vijayaraghavan, Ajey M. Joshi et al. ``In-situ sensing, process monitoring and machine control in Laser Powder Bed Fusion: A review." Additive Manufacturing 45 (2021): 102058.

\bibitem{Pandiyan}Pandiyan, V., Drissi-Daoudi, R., Shevchik, S., Masinelli, G., Le-Quang, T., Logé, R., Wasmer, K. (2021). ``Semi-supervised Monitoring of Laser powder bed fusion process based on acoustic emissions". Virtual and Physical Prototyping, 16(4), 481-497.

\bibitem{Lott}Lott, Philipp, Henrich Schleifenbaum, Wilhelm Meiners, Konrad Wissenbach, Christian Hinke, and Jan Bültmann. ``Design of an optical system for the in situ process monitoring of selective laser melting (SLM)." Physics Procedia 12 (2011): 683-690.

\bibitem{Liu}Liu, Tao, et al. ``In-situ infrared thermographic inspection for local powder layer thickness measurement in laser powder bed fusion." Additive Manufacturing 55 (2022): 102873.

\bibitem{Lough2}Lough, Cody S., Xin Wang, Christopher C. Smith, Robert G. Landers, Douglas A. Bristow, James A. Drallmeier, Ben Brown, and Edward C. Kinzel. ``Correlation of SWIR imaging with LPBF 304L stainless steel part properties." Additive Manufacturing 35 (2020): 101359.

\bibitem{Renken2}Renken, Volker, Axel von Freyberg, Kevin Schünemann, Felix Pastors, and Andreas Fischer. ``In-process closed-loop control for stabilising the melt pool temperature in selective laser melting." Progress in Additive Manufacturing 4, no. 4 (2019): 411-421.


\bibitem{Kanko}Kanko, Jordan A., Allison P. Sibley, and James M. Fraser. ``In situ morphology-based defect detection of selective laser melting through inline coherent imaging." Journal of Materials Processing Technology 231 (2016): 488-500.

\bibitem{Cheng}Cheng, Bo, James Lydon, Kenneth Cooper, Vernon Cole, Paul Northrop, and Kevin Chou. ``Infrared thermal imaging for melt pool analysis in SLM: a feasibility investigation." Virtual and Physical Prototyping 13, no. 1 (2018): 8-13.

\bibitem{Cheng2}Cheng, Bo, James Lydon, Kenneth Cooper, Vernon Cole, Paul Northrop, and Kevin Chou. ``Melt pool sensing and size analysis in laser powder-bed metal additive manufacturing." Journal of Manufacturing Processes 32 (2018): 744-753.

\bibitem{Mazzoleni}Mazzoleni, Luca, Ali Gokhan Demir, Leonardo Caprio, Matteo Pacher, and Barbara Previtali. ``Real-time observation of melt pool in selective laser melting: Spatial, temporal, and wavelength resolution criteria." IEEE Transactions on Instrumentation and Measurement 69, no. 4 (2019): 1179-1190.

\bibitem{Zhong}Zhong, Qi, Xiaoyong Tian, Xiaokang Huang, Cunbao Huo, and Dichen Li. "Using feedback control of thermal history to improve quality consistency of parts fabricated via large-scale powder bed fusion." Additive Manufacturing 42 (2021): 101986.

\bibitem{Clijsters}Clijsters, Stijn, Tom Craeghs, Sam Buls, Karolien Kempen, and J-P. Kruth. "In situ quality control of the selective laser melting process using a high-speed, real-time melt pool monitoring system." The International Journal of Advanced Manufacturing Technology 75, no. 5-8 (2014): 1089-1101.

\bibitem{Krauss}Krauss, Harald, Thomas Zeugner, and Michael F. Zaeh. ``Layerwise monitoring of the selective laser melting process by thermography." Physics Procedia 56 (2014): 64-71.

\bibitem{Krauss2}Krauss, Harald, Thomas Zeugner, and Michael F. Zaeh. ``Thermographic process monitoring in powderbed based additive manufacturing." In AIP Conference Proceedings, vol. 1650, no. 1, pp. 177-183. American Institute of Physics, 2015.


\bibitem{Sammons3}Sammons, Patrick M., Douglas A. Bristow, and Robert G. Landers. ``Two-dimensional modeling and system identification of the laser metal deposition process." Journal of Dynamic Systems, Measurement, and Control 141.2 (2019): 021012.

\bibitem{Gegel}Gegel, Michelle L., Douglas A. Bristow, and Robert G. Landers. "A loop-shaping method for frequency-based design of layer-to-layer control for laser metal deposition." In 2020 American Control Conference (ACC), pp. 487-491. IEEE, 2020.

\bibitem{Hoelzle4}Lim, Ingyu, David J. Hoelzle, and Kira L. Barton. ``A multi-objective iterative learning control approach for additive manufacturing applications." Control Engineering Practice 64 (2017): 74-87.

\bibitem{Hoelzle6} Wang, Zhi, Christopher P. Pannier, Kira Barton, and David J. Hoelzle. ``Application of robust monotonically convergent spatial iterative learning control to microscale additive manufacturing." Mechatronics 56 (2018): 157-165.

\bibitem{Hoelzle10} Afkhami, Zahra, David J. Hoelzle, and Kira Barton. ``Robust higher-order spatial iterative learning control for additive manufacturing systems." IEEE Transactions on Control Systems Technology (2023).



\bibitem{Guo}Guo, Yijie, Joost Peters, Tom Oomen, and Sandipan Mishra. ``Control-oriented models for ink-jet 3D printing." Mechatronics 56 (2018): 211-219.


\bibitem{Inyang3}Inyang-Udoh, Uduak, and Sandipan Mishra. ``A Physics-Guided Neural Network Dynamical Model for Droplet-Based Additive Manufacturing." IEEE Transactions on Control Systems Technology (2021).

\bibitem{Yavari3} Yavari, Reza, Ziyad Smoqi, Alex Riensche, Ben Bevans, Humaun Kobir, Heimdall Mendoza, Hyeyun Song, Kevin Cole, and Prahalada Rao. ``Part-scale thermal simulation of laser powder bed fusion using graph theory: Effect of thermal history on porosity, microstructure evolution, and recoater crash." Materials \& Design 204 (2021): 109685.

\bibitem{Spector}Spector, Michael JB, Yijie Guo, Souvik Roy, Max O. Bloomfield, Antoinette Maniatty, and Sandipan Mishra. ``Passivity-based iterative learning control design for selective laser melting." In 2018 Annual American Control Conference (ACC), pp. 5618-5625. IEEE, 2018.

\bibitem{Zhang2} Zhang, Zhidong, et al. ``3-Dimensional heat transfer modeling for laser powder-bed fusion additive manufacturing with volumetric heat sources based on varied thermal conductivity and absorptivity." Optics \& Laser Technology 109 (2019): 297-312.

\bibitem{Grasso} Grasso, Marco, and Bianca Maria Colosimo. ``Process defects and in situ monitoring methods in metal powder bed fusion: a review." Measurement Science and Technology 28.4 (2017): 044005.

\bibitem{H_Peng} Peng, Hao, Morteza Ghasri-Khouzani, Shan Gong, Ross Attardo, Pierre Ostiguy, Bernice Aboud Gatrell, Joseph Budzinski et al. ``Fast prediction of thermal distortion in metal powder bed fusion additive manufacturing: Part 1, a thermal circuit network model." Additive Manufacturing 22 (2018): 852-868.

\bibitem{R_Paul} Paul, Ratnadeep, Sam Anand, and Frank Gerner. ``Effect of thermal deformation on part errors in metal powder based additive manufacturing processes." Journal of manufacturing science and Engineering 136, no. 3 (2014).

\bibitem{Berumen}Berumen, Sebastian, Florian Bechmann, Stefan Lindner, Jean-Pierre Kruth, and Tom Craeghs. ``Quality control of laser-and powder bed-based Additive Manufacturing (AM) technologies." Physics procedia 5 (2010): 617-622. 

\bibitem{Wang}Wang, Xin, Cody S. Lough, Douglas A. Bristow, Robert G. Landers, and Edward C. Kinzel. ``A Layer-to-layer Control-Oriented Model for Selective Laser Melting." In 2020 American Control Conference (ACC), pp. 481-486. IEEE, 2020.

\bibitem{Ludyk}Ludyk, Günter. ``Stability of time-variant discrete-time systems''. Vol. 5. Springer-Verlag, 2013.

\bibitem{Wieberg}Wieberg, Donald M. ``Theory and Problems of State Space and Linear System''. McGraw-Hill, 1971.

\bibitem{Katsuhiko}Katsuhiko, O. (2010). ``Modern control engineering''

\bibitem{Leissner}Leissner, Patrik, Svante Gunnarsson, and Mikael Norrlof. ``Some controllability aspects for iterative learning control". Asian Journal of Control 21, no. 3 (2019): 1057-1063.

\bibitem{America Makes}America Makes, ``4039 Development \& demonstration of open-source protocols for powder bed fusion AM"
\end{thebibliography}
\end{document}